\documentclass[10pt,journal,compsoc]{IEEEtran}
\ifCLASSOPTIONcompsoc
  \usepackage[nocompress]{cite}
\else
  \usepackage{cite}
\fi
\ifCLASSINFOpdf
  \usepackage[pdftex]{graphicx}
  \graphicspath{{figures/}{pictures/}{images/}{./}}
  \DeclareGraphicsExtensions{.pdf,.jpeg,.jpg,.png}
\else
\fi
\usepackage{amsmath}
\usepackage{algorithmic}
\usepackage{mhchem}

\usepackage{enumitem} 

\usepackage{xcolor} 

\usepackage{caption}
\usepackage{subcaption}

\usepackage{breqn} 

\usepackage{mathrsfs} 

\usepackage{amsfonts} 
\usepackage{amssymb} 

\usepackage{hyperref} 
\hypersetup{
    colorlinks=true,
    linkcolor=blue,
    filecolor=magenta,      
    urlcolor=blue, 
}

\usepackage{textcomp} 

\usepackage[normalem]{ulem}

\newcommand{\remark}[1]{\textcolor{red}{#1}}

\definecolor{custom_color}{rgb}{0.6,0.2,0.6}

\newcommand{\removed}[1]{\textcolor{red}{\sout{#1}}}

\def \cleanversion{} 
\ifx\cleanversion\undefined
\else
  
  \renewcommand{\remark}{}
  \renewcommand{\removed}[1]{\iffalse#1\fi}
\fi

\newcommand{\vect}[1]{\boldsymbol{#1}} 

\begin{document}
%
\title{Geometry-Driven Detection, Tracking and Visual Analysis of Viscous and Gravitational Fingers}

\author{Jiayi~Xu,
        Soumya~Dutta, 
        Wenbin~He, 
        Joachim~Moortgat,
        and~Han-Wei~Shen
\IEEEcompsocitemizethanks{\IEEEcompsocthanksitem J. Xu is with the Department of Computer Science and Engineering, The Ohio State University, Columbus, OH, 43210.\protect\\
E-mail: xu.2205@osu.edu
\IEEEcompsocthanksitem S. Dutta is with the Data Science at Scale Team, Los Alamos National Laboratory, Los Alamos, NM 87545.\protect\\
E-mail: sdutta@lanl.gov
\IEEEcompsocthanksitem W. He is with the Department of Computer Science and Engineering, The Ohio State University, Columbus, OH, 43210.\protect\\
E-mail: he.495@osu.edu
\IEEEcompsocthanksitem J. Moortgat is with the School of Earth Sciences, The Ohio State University, Columbus, OH, 43210.\protect\\
E-mail: moortgat.1@osu.edu
\IEEEcompsocthanksitem H.-W. Shen is with the Department of Computer Science and Engineering, The Ohio State University, Columbus, OH, 43210.\protect\\
E-mail: shen.94@osu.edu}
\thanks{Manuscript received xxx xx, 201x; revised xxx xx, 201x.}}

%
%

\markboth{\tiny{\textcopyright~\textit{\lowercase{2020 \uppercase{IEEE}. \uppercase{T}his is the author`s version of the article that has been published in \uppercase{IEEE} \uppercase{T}ransactions on \uppercase{V}isualization and \uppercase{C}omputer \uppercase{G}raphics. \uppercase{T}he final version of this record is available at: \href{https://doi.org/10.1109/TVCG.2020.3017568}{10.1109/\uppercase{TVCG}.2020.3017568}}}}}{Xu \MakeLowercase{\textit{et al.}}: Geometry-Driven Detection, Tracking and Visual Analysis of Viscous and Gravitational Fingers}

%



\IEEEtitleabstractindextext{%
\begin{abstract}
Viscous and gravitational flow instabilities cause a displacement front to break up into finger-like fluids. The detection and evolutionary analysis of these fingering instabilities are critical in multiple scientific disciplines such as fluid mechanics and hydrogeology. However, previous detection methods of the viscous and gravitational fingers are based on density thresholding, which provides limited geometric information of the fingers. The geometric structures of fingers and their evolution are important yet little studied in the literature. In this work, we explore the geometric detection and evolution of the fingers in detail to elucidate the dynamics of the instability. We propose a ridge voxel detection method to guide the extraction of finger cores from three-dimensional (3D) scalar fields. After skeletonizing finger cores into skeletons, we design a spanning tree based approach to capture how fingers branch spatially from the finger skeletons. Finally, we devise a novel geometric-glyph augmented tracking graph to study how the fingers and their branches grow, merge, and split over time. Feedback from earth scientists demonstrates the usefulness of our approach to performing spatio-temporal geometric analyses of fingers. 
\end{abstract}




}

\maketitle

\IEEEdisplaynontitleabstractindextext

\ifCLASSOPTIONpeerreview
\begin{center} \bfseries EDICS Category: 3-BBND \end{center}
\fi
%
\IEEEpeerreviewmaketitle

\IEEEraisesectionheading{\section{Introduction}\label{sec:introduction}}

%
%
%
%
\IEEEPARstart{I}{n} the context of fluid flow in subsurface porous media (e.g., rock formations), \textit{fingering} refers to flow instabilities when either an invading fluid has a much lower viscosity than the displaced fluid (i.e., viscous fingering), or when a denser fluid flows on top of a lighter fluid (i.e., gravitational fingering). Fingering instabilities lead to a highly non-linear and complex evolution of the displacement front between different fluids \cite{moortgat2016viscous, amooie2018solutal}. Understanding and tracking flow instabilities is critical in a variety of scientific fields including fluid mechanics \cite{homsy1987viscous, amooie2017hydrothermodynamic}, computational fluid dynamics (CFD) \cite{luciani2018details}, and hydrogeology \cite{skauge20122, moortgat2016viscous}. Fingering is generally detrimental when the objective is to displace a viscous fluid (e.g., oil recovery through waterflooding) but can be beneficial when the aim is to mix two fluids, for instance, in the sequestration of carbon dioxide (\ce{CO2}) \cite{soltanian2016critical, amooie2017mixing} in deep water-saturated formations. 

In this work, we focus on the latter application, in which gravitational fingering helps to mix dissolved \ce{CO2} throughout the aquifer. This, in turn, helps to guarantee the storage permanence of \ce{CO2} \cite{soltanian2017dissolution, amooie2018solutal}. We illustrate the challenges and high-level motivations for detection and visualization of the fingering process from two perspectives: (1) domain-specific requirements obtained from an expert, and (2) limitations of previous evolutionary analyses for viscous and gravitational fingers. Below we expand on each of these. 

We involved an expert in Earth Sciences to help us comprehend the domain-specific requirements and challenges of this work. Flow instabilities, whether in the subsurface or space, have recently been found to obey specific universal scaling laws (e.g., \cite{soltanian2017dissolution, amooie2018solutal}); such scaling laws can be used to estimate the severity and evolution of flow instabilities for different sets of conditions, without having to redo high-resolution and thus computationally expensive simulations. To reveal those scaling laws, the geometric analysis of finger formations is critical, i.e., 
the connectivity between fingers in space and the onset, growth, merging, and splitting of fingers over time. 
\remark{However, it is difficult for the earth scientist to use standard visualization tools to track and quantify these 3D features. }

Detecting fingers is challenging because fingers are unstable structures in the fluids, and result from complex fluid interactions. Different detection techniques have been proposed~\cite{skauge20122, fu2013pattern, aldrich2016viscous, favelier2016visualizing, lukasczyk2017viscous, luciani2018details}. 
\remark{The existing detection techniques mostly use a density (or concentration) thresholding based method to detect the complete volume of fingers. However, the features detected by such density thresholding capture limited information on the internal geometric structures of fingers. }

In this study, we propose a geometry-driven solution to satisfy the requirements of scientists and analyze geometric structures of fingers. Guided by a ridge voxel detection method, we first extract finger cores from the data of 3D density fields. Based on the extracted finger cores, we obtain the complete volume of fingers, and produce finger skeletons to acquire the overall geometric features of fingers in space. We then propose a spanning tree based algorithm to construct finger branches from finger skeletons. We visualize fingers and their branches as geometric glyphs, and nest the glyphs into a tracking graph so that we can track and compare fingers and branches efficiently and effectively. Hence, our contributions are twofold: 
\begin{enumerate} 
\item We propose voxel-based ridge detection, which is inspired by Steger's pixel-based method \cite{steger1998unbiased}, to guide the extraction of finger cores on 3D scalar fields. 
Furthermore, we provide a spanning tree based heuristic algorithm for the construction of finger branches from finger skeletons. 
\item We offer an interactive visual analytics system that allows efficient and effective exploration of fingers over space and time with minimized occlusion. Our system incorporates a novel geometric-glyph augmented tracking graph that reveals the temporal evolution of the fingers and their branches. 
\end{enumerate}

\section{Application Background and Requirements} \label{sect:background}
In this section, we elaborate on the concepts for the formation of fingers, discuss the domain-specific requirements, and summarize the limitations of previous detection and visualizations of viscous and gravitational fingers. 

\subsection{Viscous and Gravitational Fingers}
Viscous and gravitational flow instabilities in porous media result in finger-like features. The \textit{fingering} phenomenon refers to the formation and evolution of such fingers. The fingering instabilities are triggered by adverse mobility or density ratios between displacing and displaced fluids \cite{moortgat2016viscous}. \textit{Viscous fingering} is caused by viscosity contrasts between fluids: when a less viscous fluid is injected into a more viscous medium, the less viscous fluid tends to penetrate through the more viscous fluid to form elongated finger-like structures \cite{homsy1987viscous}. 
\textit{Gravitational fingering} is caused by contrasts in density between fluids: when a denser fluid resides on top of a less dense fluid, the interface may become unstable. The denser fluid vertically penetrates the lighter fluid to form fingers, while the lighter fluid rises buoyantly. 


\textbf{Data description:} The fingering datasets are generated from simulations of injecting carbon dioxide (\ce{CO2}) from the top of a water-saturated reservoir. The gravitational fingering helps to mix injected \ce{CO2} throughout the aquifer. When \ce{CO2} dissolves in water at the top, it locally increases the water density, which is prone to gravitational instabilities. The \ce{CO2}-waterfront becomes unstable and leads to fingering of \ce{CO2}-enriched, denser, water downwards with the upwelling of fresh lighter water. The space of the data is defined within
a 3D rectilinear grid ($90 \times 90 \times 100$). The density is defined at the center of every grid cell; in other words, the density data generated by the simulation are cell-centered. Each simulation has more than one hundred timesteps.


\subsection{Scientist Requirements} \label{sect:expert_requirements}
We discuss the application-specific requirements that were identified by a domain expert who specializes in Earth Sciences. This earth scientist has ten years of experience in researching fluid injection processes and the associated viscous and gravitational fingering instabilities. We generalize three requirements in the following. 


\subsubsection{Requirement One (\textbf{R1}): Identification of Geometric Features}
Visualizing and quantifying the geometric features of fingers provide insights into the analyses of bimolecular reaction \cite{de2014filamentary} and physical flow regimes of, e.g., enhanced or suppressed mixing rates \cite{amooie2017mixing, amooie2017hydrothermodynamic}. Moreover, the geometrical features, including widths \cite{soltanian2016critical} of fingers and locations \cite{soltanian2016critical} of fingertips, have been used to analyze scaling behaviors. However, the vast majority of studies \cite{de2014filamentary, soltanian2016critical} for geometric analysis of fingers are in 2D and often that has been done essentially by hand and visual inspection.

\remark{From our discussion with the earth scientist, two important geometric finger characteristics were identified: branching (\textbf{R1.1}) and height (\textbf{R1.2}). The earth scientist also explained that these two characteristics have the potential for identifying new scaling laws.} Specifically, the branching is critical to scientists in understanding flow instabilities~\cite{buka1987stability, buka1987viscous, hinrichsen1989self, li2009control}. Due to the gravity, fingers stretch vertically; based on the height, we can estimate the vertical speed of growth of fingers to analyze the stretching process. In this paper, we define \textit{height (persistence) of a finger} as the height difference between the finger root and the fingertip. The \textit{finger root} is the highest part of the finger and, generally, is where the $\ce{CO2}$ is injected. The \textit{fingertip} is the deepest point that the finger can reach in the reservoir, and usually has a low-density value. 
\remark{The location of the finger root and fingertip is illustrated in Fig.~\ref{fig:simple-finger}b. }
Similarly, \textit{height (persistence) of a finger branch} is the height difference between the highest point and deepest point on the branch. 

\subsubsection{Requirement Two (\textbf{R2}): Spatial Exploration}
During our interactions, the scientist further mentioned that he was interested in how to visualize the 4D (3D in space plus time) simulations of the fluid injection. In his day to day studies, the expert often visually analyzes the fingers using visualization tools such as ParaView \cite{ayachit2015paraview}, VisIt \cite{childs2012visit}, and Tecplot \cite{tecplot2018}. However, the expert informed us that the current visualization techniques that he used were not ideal. Tracking and quantifying the 3D geometry of these ramified structures in both space and time is virtually impossible with those methods. For example, some domain tools visualize fingers as hollow sheets rather than dense columns; also, the three-dimensional fingers which were visualized by standard visualization methods such as volume rendering and isosurfaces suffered from the occlusion problem. Thus, the expert usually had to cut multiple cross-sections of those fingers to analyze the formation and internal structures of fingers rigorously. 

The scientist motivated us to develop visualizations to explore fingers in space effectively and efficiently. Specifically, the visualizations were required to address how hundreds of distributed fingers grow vertically (\textbf{R2.1}) and spread horizontally (\textbf{R2.2}) with minimal occlusion (\textbf{R2.3}). 

\subsubsection{Requirement Three (\textbf{R3}): Temporal Exploration}
The earth scientist was specifically interested to know how the fingers shield \remark{(e.g., one finger shields the other finger from growing further \cite{homsy1987viscous})} and merge at certain timesteps and then predominantly split into new smaller fingers at other timesteps geometrically (\textbf{R3.1}). The expert thought that interactive space-time diagrams of the fingers, which can effectively present the finger-specific evolution events such as merging, splitting, and branching of fingers, would be extremely valuable.

\subsection{Limitations of Existing Works for Fingers}
In this section, we review the methods used previously to detect and visualize viscous and gravitational fingers, and highlight the limitations of the existing methods. 

\subsubsection{Limitations of Current Detection Methods for Fingers}
Since the formation of fingers is extremely non-linear, accurate detection of fingers is a non-trivial task. 
\remark{To extract fingers from the density (or concentration) scalar fields, previous researchers typically used thresholding on the density value to extract high-density regions. Then, they interpreted the connected high-density regions as fingers. }
In the previous study, Skauge et al. \cite{skauge20122} and Fu et al. \cite{fu2013pattern} modeled fingers as high-density vertical lines, which were detected by a peak detection method. 
Aldrich et al. \cite{aldrich2016viscous} and Lukasczyk et al. \cite{lukasczyk2017viscous} considered each finger to be a connected component of high-density 3D regions, and detected fingers through the identification of connected components. 
\remark{
Favelier et al. \cite{favelier2016visualizing} detected fingertips first, and then segmented the input volume from the fingertips to isolate the complete fingers by using a watershed traversal method \cite{vincent1991watersheds}. }
Luciani et al. \cite{luciani2018details} studied particle data of fingering, and grouped particles with close locations and concentrations to be fingers. 


The previous detection methods \cite{skauge20122, fu2013pattern, aldrich2016viscous, favelier2016visualizing, lukasczyk2017viscous, luciani2018details} typically detect each finger as a single entity by using density thresholding. Even though the previous methods obtained good results, the density thresholding does not capture detailed geometric structures of fingers, which are essential for earth sciences (R1.1). Specifically, fingertips usually have much lower densities than the finger roots. 
\remark{If the used threshold is too high, the fingertips with low density values may not be detected. 
If the used threshold is too low, different finger branches may not be segmented because they are connected by low-density regions as shown in Fig.~\ref{fig:simple-finger}d. }




\subsubsection{Limitations of Current Tracking Graphs for Fingers}
When it comes to the visualization of time-varying fingers, none of the existing tracking visualizations focus on the geometric evolution of the fingers. The tracking graphs provided by Aldrich et al. \cite{aldrich2016viscous} display fingers at each timestep as points in a column. Aldrich et al. \cite{aldrich2016viscous} used hues to encode different fingers and linked related fingers between adjacent timesteps by curves. In recent work, Lukasczyk et al.~\cite{lukasczyk2017nested} proposed nested tracking graphs, which can depict the evolution of level-sets of density fields and visualize the evolution of fingers across multiple specified density thresholds. The lack of tracking graphs for the geometric evolution of fingers that encodes the branching information further motivates us to develop a new geometry-driven tracking graph for the analysis of the fingering process.

\section{Technical Background: Ridges and Reeb Graphs}
We offer the background of two concepts: ridge and reeb graph, which are the bases of the geometry-driven approaches applied in this paper. 

\subsection{Ridges}
\subsubsection{Ridge Definition} \label{sect:ridge_definition}
Ridges originally are structures of surface topography whose mathematical properties were studied by de Saint-Venant \cite{de1852surfaces}. The concept of $k$-dimensional ridges is generalized by Haralick \cite{haralick1983ridges}, and re-formulated by Lindeberg \cite{lindeberg1998edge} and Eberly \cite{eberly2012ridges}. Intuitively, the ridges of a smooth function are a set of curves or surfaces whose points are local maxima of the function in certain dimensions. 

A formal description of ridges is given in the following. Given a scalar field $f$: $\mathbb{R}^n \to \mathbb{R}$, we define $\vec{v_1},...,\vec{v_n}$ as the eigenvectors of the Hessian matrix $H=[\frac{\partial ^2 f}{\partial x_i \partial x_j}]$, where $x_i$ and $x_j$ represent any two axes of the coordinate system. The $n$ eigenvectors are ordered based on their corresponding eigenvalues $\lambda_1,...,\lambda_n$. Two alternative ways were proposed to order the eigenvalues. 
Lindeberg \cite{lindeberg1998edge} ordered the eigenvalues by 
$|\lambda_1| \geq ... \geq |\lambda_n|$, and 
Eberly \cite{eberly2012ridges} ordered the eigenvalues by 
$\lambda_1 \leq ... \leq \lambda_n$; \remark{the results presented in this paper are generated by using the Lindeberg's ordering \cite{lindeberg1998edge}.} The \textit{$k$-dimensional ridge} is defined as a set of points where the following two conditions are satisfied: given any point $\vect{p}$ in the set,
\begin{description} 
\item [Condition One for extreme point determination: ]
Intuitively, $\vect{p}$ is a local extreme point along the directions of the first $n-k$ eigenvectors \cite{eberly2012ridges}. Mathematically, the first-order directional derivatives at $\vect{p}$ along the first $n-k$ eigenvectors are all zeros,  
\begin{equation} \label{equa:condition_two}
D_{\vec{v_1}}f(\vect{p})=...=D_{\vec{v_{n-k}}}f(\vect{p})=0
\end{equation}
\item [Condition Two for feature type differentiation: ]
This condition differentiates ridge points from other types of extreme points. Intuitively, the scalar value at $\vect{p}$ is maximal along the first $n-k$ eigenvectors. Mathematically, the second-order directional derivatives at $\vect{p}$ along the first $n-k$ eigenvectors are all smaller than zero,  
\begin{equation} 
D_{\vec{v_1}}^2f(\vect{p}),...,D_{\vec{v_{n-k}}}^2f(\vect{p})<0
\end{equation}
Since $\lambda_i=\nabla_{\vec{v_i}}^2f(\vect{p})$, we also have
\begin{equation} 
\lambda_1,...,\lambda_{n-k}<0
\end{equation}
\end{description}
In this paper, as the space is $\mathbb{R}^3$, \textit{ridge voxels} are defined as voxels that contain $1$-dimensional ridge points, and are where the ridge lines pass through. 


\subsubsection{Ridge Detection} \label{sect:related_work_ridge_detection}

The concept of ridges has been used to detect and extract curvilinear structures in computer vision \cite{steger1998unbiased, lindeberg1998edge} and scientific visualization \cite{furst2001marching, peikert2008height}. 
\remark{In computer vision, one can regard a greyscale image as a 2D scalar field, and extract curvilinear structures such as roads \cite{steger1998unbiased} and human fingers \cite{lindeberg1998edge} based on ridge detection. }
In scientific visualization, definitions of ridges are extended to volumetric data \cite{furst2001marching, peikert2008height}. \remark{
Although the whole ridge structure is complex, ridge points are local features; namely, they are the features that satisfy the local criterion \cite{lindeberg1998edge, eberly2012ridges} consisting of the two conditions mentioned above. }
So, the locally approximated function and derivatives can detect the ridge points and approximate the entire ridge structures well \cite{steger1998unbiased, furst2001marching, peikert2008height}. When the scalar value is defined at every grid point, the marching ridge technique \cite{furst2001marching} estimates the derivatives at grid points of a given grid cell; then, the derivatives at the grid points are used to interpolate the ridge lines within the grid cell by using zero-crossing tri-linear \remark{interpolation}. 


\remark{As a scalar value is defined at the center of every grid cell, such as the pixel-based image in $\mathbb{R}^2$, Steger \cite{steger1998unbiased} constructed a Taylor polynomial for every pixel to detect the pixels that contain ridge points. }
\remark{In our application, we generalize the Steger's method from $\mathbb{R}^2$ to $\mathbb{R}^3$ to detect $3$-dimensional grid cells that contain $1$-dimensional ridge points, and explain the challenge for the ridge voxel detection in $\mathbb{R}^3$ as follows. }



Given a grid cell, we estimate the positions of the ridge points locally, and determine whether the position of any local ridge point is in or on the boundary of this grid cell. Since the function of the scalar field, $f$, is unknown and only discrete scalar values at cell centers are observed, it is hard to acquire the positions of the ridge points directly. To remedy this, Steger \cite{steger1998unbiased} approximated $f(\vect{p})$ locally by using the second-order Taylor polynomial $g(\vect{p})$ centered at every grid cell. According to Taylor's theorem, the Taylor polynomials can approximate the scalar function with low error, and it is efficient to compute the derivatives from the Taylor polynomials. Hence, the Taylor polynomials are useful for locating the ridge points. 
The second-order Taylor polynomial $g(\vect{p})$ is defined by \remark{
\begin{dmath} \label{equa:taylor-polynomial}
g(\vect{p}) = f(\vect{p_0}) + \nabla f(\vect{p_0})^\intercal (\vect{p}- \vect{p_0}) + \frac{1}{2}(\vect{p}- \vect{p_0})^\intercal H(\vect{p_0}) \cdot (\vect{p}- \vect{p_0})
\end{dmath}}
where $\vect{p_0}$ is the center point of the given grid cell. 
$\nabla f(\vect{p_0}) = (\frac{\partial f(\vect{p_0})}{\partial x_1},\frac{\partial f(\vect{p_0})}{\partial x_2},\frac{\partial f(\vect{p_0})}{\partial x_3})^\intercal$ is the estimated gradient vector at $\vect{p_0}$ \remark{by using the central difference}. 
$H(\vect{p_0})=[\frac{\partial ^2 f(\vect{p_0})}{\partial x_i \partial x_j}]$ is the estimated Hessian matrix at $\vect{p_0}$ \remark{by using the central difference}. 
Given $g(\vect{p})$ in Equation~\ref{equa:taylor-polynomial}, we can approximate the gradient vector at $\vect{p}$ by 
\begin{dmath} \label{equa:jacobian_matrix_p}
\nabla f(\vect{p}) \approx \nabla g(\vect{p}) = (\frac{\partial g(\vect{p})}{\partial x_1},\frac{\partial g(\vect{p})}{\partial x_2},\frac{\partial g(\vect{p})}{\partial x_3})^\intercal = 
\nabla f(\vect{p_0}) + H(\vect{p_0}) \cdot (\vect{p}- \vect{p_0})
\end{dmath}
The first-order directional derivatives are approximated by
\begin{dmath} \label{equa:directional_derivative}
D_{\vec{v_{i}}}f(\vect{p}) \approx
D_{\vec{v_{i}}}g(\vect{p}) 
= \vec{v_i}^\intercal \cdot \nabla g(\vect{p})
=\vec{v_i}^\intercal \cdot [\nabla f(\vect{p_0}) + H(\vect{p_0}) \cdot (\vect{p} - \vect{p_0})]
\end{dmath}
Also, the Hessian matrix is estimated by
\begin{equation} \label{equa:hessian_matrix_p}
H(\vect{p}) \approx [\frac{\partial ^2 g(\vect{p})}{\partial x_i \partial x_j}] = H(\vect{p_0}) 
\end{equation}

To locate local ridge points efficiently, Steger's method \cite{steger1998unbiased} has one additional assumption: the local ridge points should lie on the line that passes the center $\vect{p_0}$ of the grid cell and is at the direction of $\vec{v_1}$. 
\remark{Based on this assumption, whether any point $\vect{p}$ on this line satisfies Condition One, i.e., the extreme point determination, can be efficiently identified. }
Second, for Condition Two (i.e., the feature type differentiation), one examines whether the first two eigenvalues of $H(\vect{p})$ are all smaller than zero or not to guarantee the extreme point $\vect{p}$ is a ridge point. 
Third, if the position of any identified local ridge point is within or on the boundary of the grid cell, the grid cell is claimed to contain ridge points. 
Although the abovementioned assumption gets success in $\mathbb{R}^2$, a challenge is that this assumption brings more restriction to the detection of ridge points within grid cells in $\mathbb{R}^3$ and may lead to missing ridge voxels. Hence, in order to solve this challenge, our approach derives inequalities solely from the basic ridge criterion \cite{lindeberg1998edge, eberly2012ridges} to detect ridge voxels without additional assumptions for the locations of ridge points.

\subsection{Reeb Graphs}\label{sect:reeb_graph}

The \textit{Reeb graph} obtains the topology of a compact manifold by following the evolution of the level-sets of a scalar function defined on the manifold. Nodes in the Reeb graph represent critical points of the scalar function, and edges correspond to connections between critical points \cite{cornea2007curve, biasotti2008reeb}. 

\textbf{Skeletonization: } The Reeb graph based skeletonization is one of the standard skeleton-extraction approaches \cite{cornea2007curve}. When we consider the height function of an object, the \remark{corresponding} Reeb graph can capture topological features of the object. The corresponding Reeb graph is not a skeleton; however, Lazarus and Verroust \cite{lazarus1999level} embedded the Reeb graph into the space to get the skeleton of the original object which also can indicate the height of the object. If an object has no holes inside, the contour tree \cite{van1997contour, tierny2018topology}, as a special instance of the Reeb graph, is sufficient for the skeletonization of the object. 

\textbf{Trim: }
To trim Reeb graphs, Bauer et al. \cite{bauer2014measuring} suggested removing features with persistence smaller than a threshold. 
Carr et al. \cite{carr2010flexible} suggested pruning away small leaves. In addition to short leaves, Doraiswamy and Natarajan \cite{doraiswamy2012output} also suggested removing short cycles in Reeb graphs.

\begin{figure*}[tb]
  \centering 
  \includegraphics[width=\linewidth]{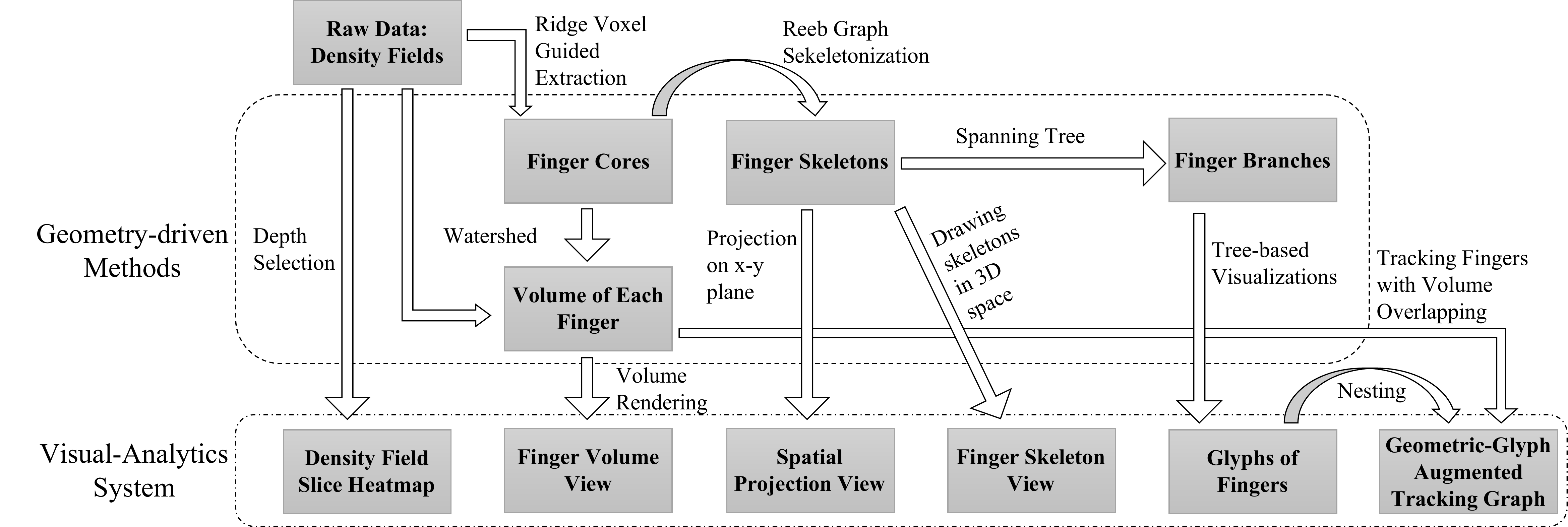}
  \caption{We present a schematic diagram of our geometry-driven approaches for viscous and gravitational fingering. }
  \label{fig:flow-chart}
\end{figure*}


\section{Approach Overview}

In this work, we focus on the detection of geometric structures of fingers (R1) and the spatio-temporal visualizations of the fingers with minimized occlusion (R2 and R3). Fig.~\ref{fig:flow-chart}, a schematic diagram, presents the pipeline of our solution. From the density fields, we apply a ridge voxel guided detection method to extract finger cores (Sect.~\ref{method:ridge_voxel}). We expand the finger cores to obtain the complete volume of fingers (Sect.~\ref{sect:segmentation}), and display the volume of fingers in the spatial domain by using volume rendering (Sect.~\ref{sect:spatial_visualization_selected_finger}). We skeletonize the finger cores into finger skeletons by using a Reeb graph based skeletonization (Sect.~\ref{method:reeb_graph}), and draw the finger skeletons in 3D space to provide the overall geometric structures of fingers (Sect.~\ref{sect:spatial_visualization_selected_finger}). Also, we project finger skeletons onto a plane to observe spatially relative positions between fingers (Sect.~\ref{vis:projection}). 
We construct finger branches from the finger skeletons \remark{by} using a variant of spanning tree method (Sect.~\ref{method:branch_hierarchy}). 
Given the branches of fingers, we adopt various tree-based visualization methods to represent different aspects of finger branches (Sect.~\ref{vis:linear_glyph} and \ref{vis:treemap}). Finally, we develop a geometric-glyph augmented tacking graph to study how the fingers evolve geometrically (Sect.~\ref{sect:tracking_graph}). On the tacking graph, we link temporally related fingers (Sect.~\ref{vis:link_encoding}) and minimize link crossings (Sect.~\ref{vis:link_crossing}). We also design interactions to identify the change of branches over time (Sect.~\ref{vis:interaction}). 






\section{Geometry-Driven Detection of Viscous and Gravitational Fingering}

In this section, we present the geometry-driven detection technique for the viscous and gravitational fingering process in detail. Since the fingering process is complex and non-linear, precise descriptors for the fingers are unavailable in the earth science domain. In order to develop a reliable technique for extracting fingers, we model the central regions of the fingers as ridges based on the diffusive process of fingers. 
In our work, the finger cores include both the ridge regions and \remark{the volume near the ridges} to capture the connections among neighboring finger branches robustly. 
The complete finger volume consists of the finger cores and the connected lower-density regions covering the finger cores. To capture the geometric features of fingers (R1), we skeletonize finger cores and construct finger branches.  


\begin{figure}[htb]
\centering

  \includegraphics[width=\columnwidth]{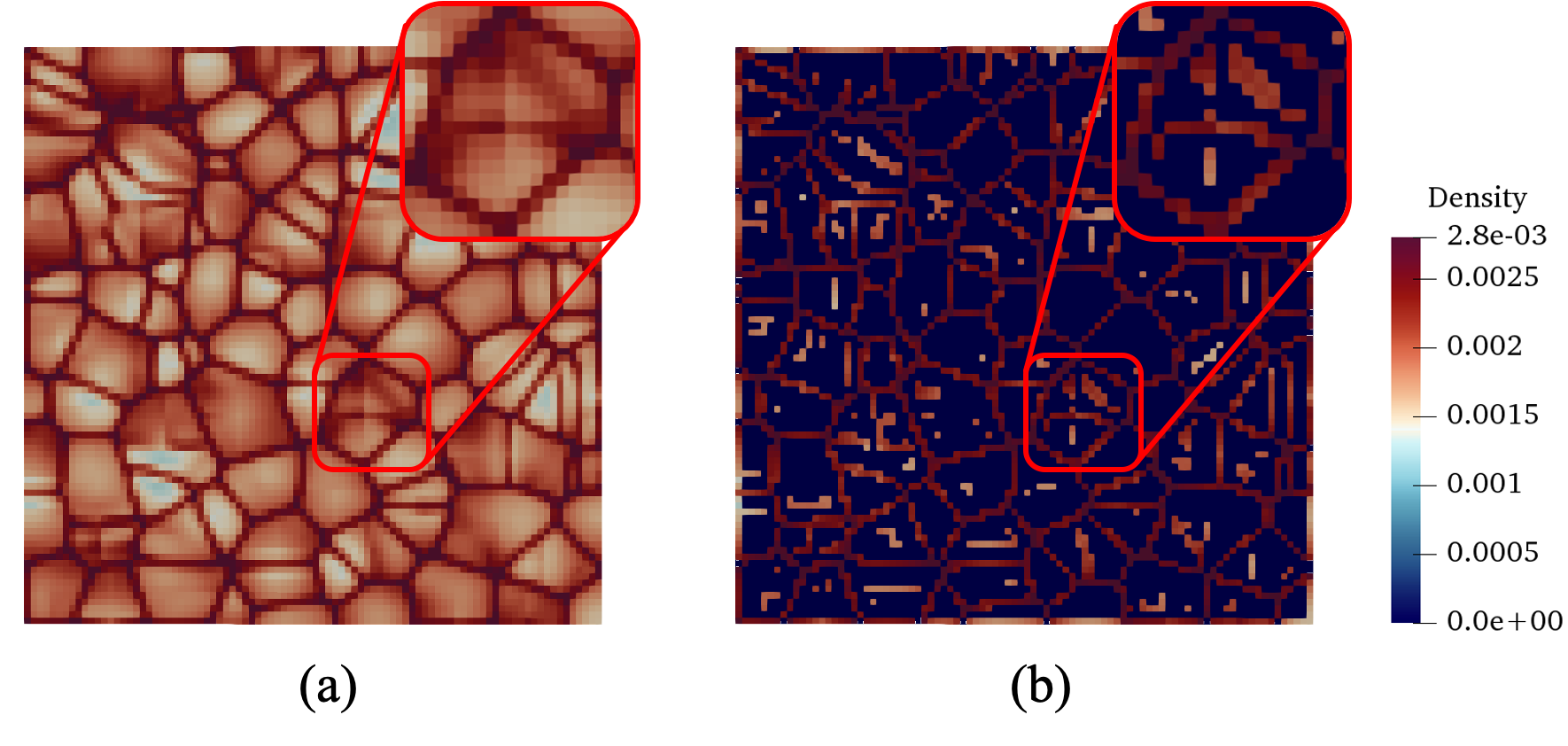}

  \caption{(a) We show the top view of the 3D density field. \remark{(b) We show the top view of the extracted ridge voxels by setting the density values of other voxels to be zero.} As the same places are highlighted with a red box in (a) and (b), the ridge voxels capture high-density hexagonal cells in the top grid cells and high-density curvilinear structures inside the hexagonal cells. }

  \label{fig:method-top-surface}
\end{figure}

\begin{figure*}[tb]
  \centering 
  \includegraphics[width=\linewidth]{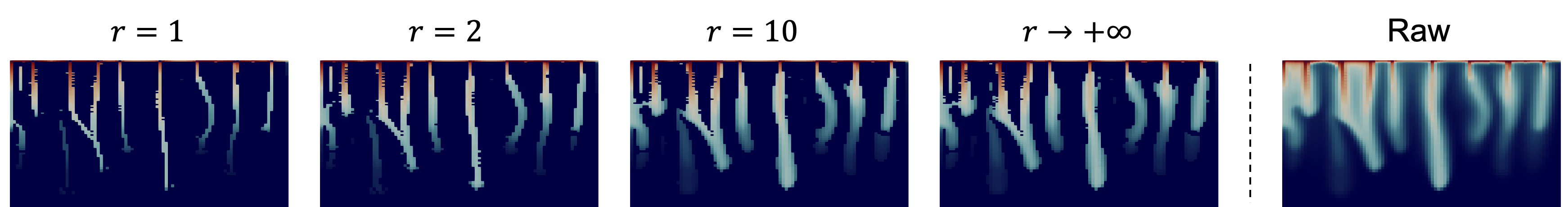}
  \caption{We display, on the left side, the sensitivity of the extracted finger cores about different $r$ values by comparing with, on the right side, the corresponding raw density field. We observe that the finger cores generally become more connected and thicker as increasing $r$. The extracted features become little changed when $r$ is larger than $10$. When $r$ becomes large enough, the voxels, whose local information cannot imply any outside ridge structures, are filtered out. }
  \label{fig:sensitivity_r}
\end{figure*}

\subsection{Ridge Voxel Guided Extraction of Finger Cores} \label{method:ridge_voxel}

The extraction of finger cores is guided by a ridge voxel detection method. According to our domain scientist, one of the causes of fingering is the diffusion process. The diffusion of \ce{CO2} is driven by compositional gradients (i.e., the difference of \ce{CO2}/water composition). Through the diffusion process, \ce{CO2} volume tends to spread out from high-density regions to low-density regions; the regions with higher densities usually form the center of finger structures. In this context, it is important to note that the ridges are regions with locally higher densities in the density fields. Therefore, the extraction of finger cores can be guided by the ridge detection technique, as discussed below. 


\subsubsection{Ridge Voxel Detection} \label{method:ridge_voxel_detection}

To obtain regions with locally higher densities, we filter the 3D density field to identify ridge voxels following the generalized ridge detection framework explained in Sect.~\ref{sect:related_work_ridge_detection}. The critical part of the detection framework is to robustly detect the ridge voxels, i.e., how to examine whether a given voxel contains any local ridge point without any additional assumptions for the locations of ridge points. In the following, we provide a solution to this. 


Intuitively, given a voxel, we first estimate the locations of extreme points surrounding the voxel. Then, we examine whether any extreme point is located inside or on the boundary of the voxel. 
Finally, we determine whether any extreme point contained by the voxel is \remark{a} ridge point to conclude that the voxel is a ridge voxel. 
Note that Condition One and Two of the ridge criteria used below are explained in Sect.~\ref{sect:ridge_definition}. 

We first estimate the positions of extreme points. After approximating the density function $f(\vect{p})$ locally by using the second-order Taylor polynomial $g(\vect{p})$, we obtain the two first-order directional derivatives, $D_{\vec{v_{1}}}g(\vect{p})$ and $D_{\vec{v_{2}}}g(\vect{p})$, through Equation~\ref{equa:directional_derivative}. We equate the two first-order directional derivatives to zeros according to Condition One (i.e., the extreme point determination) of the ridge criteria to obtain: 
\begin{equation} \label{equa:polynomial_equations}
D_{\vec{v_{i}}}g(\vect{p}) = \vec{v_i}^\intercal \cdot [\nabla f(\vect{p_0}) + H(\vect{p_0}) \cdot (\vect{p} - \vect{p_0})]
=0, {i=1,2}
\end{equation}
where $p_0(x_{1,0}, x_{2,0}, x_{3,0})$ is the center of the given voxel and $p(x_1, x_2, x_3)$ represents estimated extreme point. 
Equation~\ref{equa:polynomial_equations} constrains the coordinates of estimated extreme points, and defines the region formed by the extreme points in space. 



We examine whether any extreme point is contained by the given voxel through computing the intersection between the region formed by extreme points and the region covered by the voxel; if the intersection is not empty, the voxel contains extreme points. We define the region covered by the voxel below: 
\begin{equation} \label{squa:inequity}
x_{j,0} - \frac{s}{2} \leq x_{j} \leq x_{j,0} + \frac{s}{2},~j=1,2,3
\end{equation}
where $s$ is the length of the side of the voxel; any point $p(x_1, x_2, x_3)$ within or on the boundary of the voxel should satisfy Equation~\ref{squa:inequity}. 
We compute the intersection between the region covered by the voxel and the region formed by extreme points through plugging Equation~\ref{equa:polynomial_equations} into Equation~\ref{squa:inequity} as follows. Note that, Equation~\ref{equa:polynomial_equations} gives two polynomial equations when $i$ is replaced by $1$ and $2$ respectively, and has three unknowns ($x_{1}$, $x_{2}$, and $x_{3}$ of $\vect{p}$), hence, we can represent two of the unknowns by polynomials of the third unknown; without loss of generality, we represent $x_{2}$ and $x_{3}$ by polynomials of $x_{1}$ through transforming Equation~\ref{equa:polynomial_equations}. Thus, we can substitute $x_{2}$ and $x_{3}$ with the polynomials of $x_{1}$ for the three inequalities in Equation~\ref{squa:inequity}, which produces the intersection region represented by three inequities that are only associated with $x_{1}$. 
If the union of the three inequities of $x_{1}$ is not empty, we claim that the voxel contains extreme points. 

We examine whether any extreme point $\vect{p}$ contained by the voxel passes Condition Two, \remark{i.e.,} the feature type differentiation. 
If the first two eigenvalues of $H(\vect{p})$, estimated by Equation~\ref{equa:hessian_matrix_p}, are all smaller than zero, we declare $\vect{p}$ is a ridge point and this voxel is a ridge voxel. 

\remark{We demonstrate the results of the ridge voxel detection in Fig.~\ref{fig:method-top-surface} and Fig.~\ref{fig:sensitivity_r}.  Fig.~\ref{fig:method-top-surface} shows the top view of the extracted ridge voxels, and the ``$r=1$'' column of Fig.~\ref{fig:sensitivity_r} displays the side view of the ridge voxels.} The ridge voxels of several individual fingers are shown in Fig.~\ref{fig:simple-finger}a, Fig.~\ref{fig:medium-finger}a, and Fig.~\ref{fig:complex-finger}a.






\subsubsection{Acquisition of Additional Branch Connections} \label{sect:finger_core_extraction}

\begin{figure}[htb]
  \centering 
  \includegraphics[width=0.90\columnwidth]{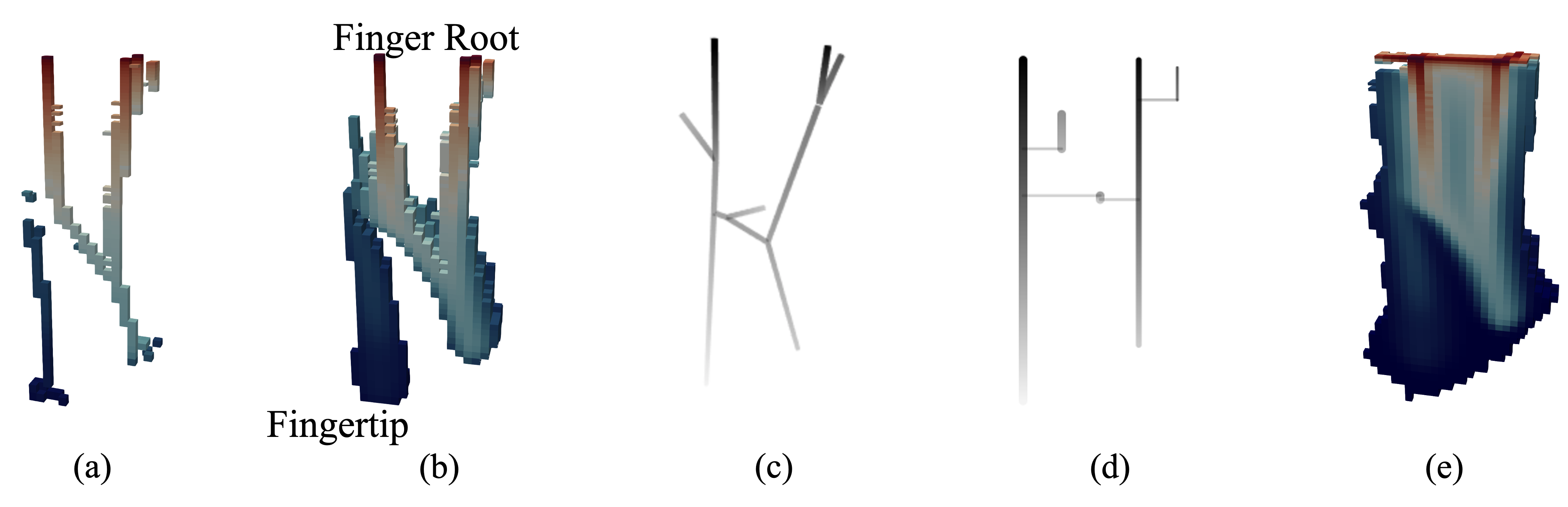}
  \caption{We display the results of our methods for a simple finger. We show five images: (a) the ridge voxels, (b) the finger core whose finger root and fingertip are indicated, (c) the finger skeleton, (d) the linear glyph of the finger, and (e) the complete finger volume. 
  }
  \label{fig:simple-finger}
\end{figure}

In addition to ridge voxels, we include voxels that are close to the ridges into the finger cores to obtain more connections between branches (R1.1). According to the study of Damon \cite{damon1999properties}, ridges do not preserve the connections between branches well enough because ridge lines can only cross at critical points. Even though the extracted ridge voxels group ridge lines when they are closer than the size of a voxel, certain finger branches are still disconnected. For instance, the fingers in Fig.~\ref{fig:simple-finger}a and Fig.~\ref{fig:medium-finger}a have disconnected branches, although the connections of branches in Fig.~\ref{fig:complex-finger}a are well-preserved. To remedy this, we acquire additional necessary branch connections by including the voxels that have ridge points nearby. 

We define the nearby region of a given voxel by a cube and examine whether that cube contains any ridge point. 
Specifically, given the center $p_0(x_{1,0},x_{2,0},x_{3,0})$ of a voxel, we create a cube centered at $\vect{p_0}$ with side length $r \cdot s$ where $r\geq 1$ and $s$ is the side length of voxel. We define the region covered by the cube:
\begin{equation} \label{squa:inequity_nearby_region}
x_{j,0} - \frac{r \cdot s}{2} \leq x_{j} \leq x_{j,0} + \frac{r \cdot s}{2},~j=1,2,3
\end{equation}
The cube fully covers the given voxel in space, and, intuitively, if $r$ is large, the cube contains more surrounding regions of the voxel. We examine whether the cube contains any ridge points by the same method in Sect.~\ref{method:ridge_voxel_detection} but replacing Equation~\ref{squa:inequity} with Equation~\ref{squa:inequity_nearby_region}. If the cube centered at the voxel contains any ridge point, we include the given voxel into finger cores. 



Scientists can control how many additional voxels are needed for the preservation of branch connections through visual inspection, as the images are shown in Fig.~\ref{fig:sensitivity_r}. In this application, we keep increasing $r$ until the results remain unchanged to maximize the acquired branch connections. The produced results are demonstrated in  Fig.~\ref{fig:simple-finger}b, Fig.~\ref{fig:medium-finger}b, and Fig.~\ref{fig:complex-finger}b respectively. The limitation of using $r$ is that it makes the branch connections sensitive to the additional parameter, $r$, which is demonstrated in Fig.~\ref{fig:sensitivity_r}. However, a benefit is that by controlling the value of $r$, scientists can flexibly study other similar datasets without changing the finger core detection algorithm. 

\subsection{Construction of Geometric Structures of Fingers}

In this section, we construct geometric structures of fingers (R1). We first obtain finger skeletons from finger cores. Finger branches are constructed from finger skeletons then. We further trim finger skeletons by removing short branches and cycles. 

\subsubsection{Reeb Graph Based Skeletonization} \label{method:reeb_graph}

\begin{figure}[htb]
  \centering 
  \includegraphics[width=0.80\columnwidth]{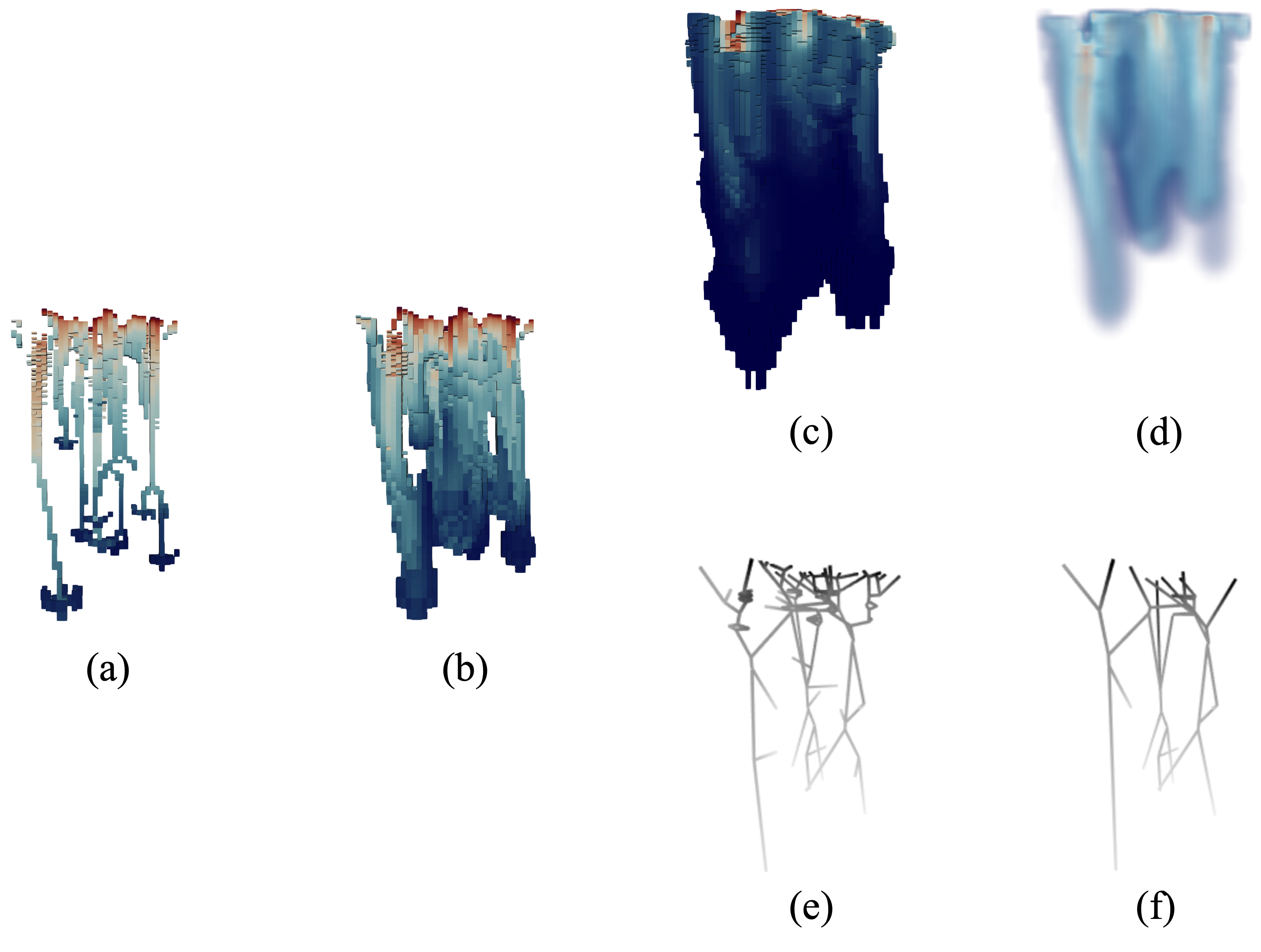}
  \caption{We display a  complex finger including (a) the ridge voxels, (b) the finger core, (c) the complete finger volume, (d) the volume rendering image of (c), (e) the finger skeleton extracted from (b), and (f) the trimmed finger skeleton. }
  \label{fig:complex-finger}
\end{figure}

We use the Reeb graph based method \cite{cornea2007curve, biasotti2008reeb, lazarus1999level, lukasczyk2017viscous, gueunet2019reeb} to obtain finger skeletons. In computational geometry, Reeb graphs are well-known for their ability to show the skeletons of a 3D object \cite{cornea2007curve} and preserve the branching and the height of a 3D object accurately (R1). Contour tree \cite{van1997contour} is an alternative approach for shape skeletonization. However, in our application, voxels that do not belong to finger cores are removed, which may leave holes in the volume of finger cores (e.g., Fig.~\ref{fig:complex-finger}b) and result in torus-like structures. 
Hence, the contour tree based method is not appropriate for this application, although it is more efficient than the Reeb graph based method. 

\begin{figure}[htb]
  \centering 
  \includegraphics[width=0.90\columnwidth]{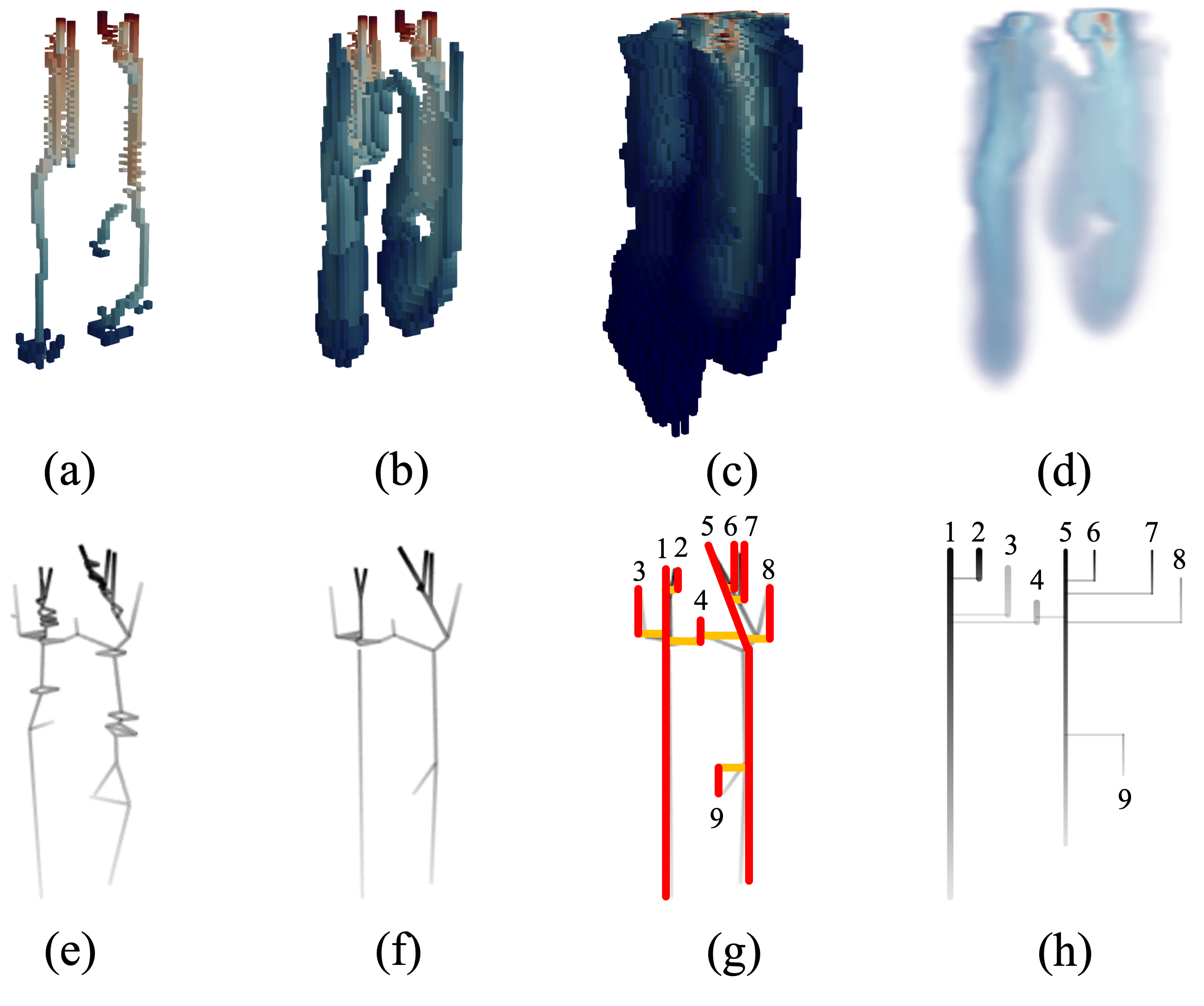}
  \caption{We display a complex finger with (a) the ridge voxels, (b) the finger core, (c) the complete finger volume, (d) the volume rendering image of (c), (e) the finger skeleton extracted from (b), (f) the trimmed finger skeleton, (g) the constructed branches from (f), and (h) the linear glyph that is based on (g). In (g), the branches are represented by red lines, and the connections between branches are represented by orange lines. In (g) and (h), the same numbers label corresponding branches. }
  \label{fig:medium-finger}
\end{figure}

The skeleton of a simple finger is shown in Fig.~\ref{fig:simple-finger}c. Complex examples are shown in (e) of both Fig.~\ref{fig:medium-finger} and Fig.~\ref{fig:complex-finger}.

\subsubsection{Spanning Tree Based Extraction of Finger Branches} \label{method:branch_hierarchy}

We extract finger branches from finger skeletons by a spanning tree based heuristic algorithm. We create finger branches explicitly for two reasons. First, we need to identify branches according to R1.1. 
Second, finger branches, as tree structures, can be visualized with minimized occlusion (R2.3). Due to the stretching process, finger branches grow downward; hence, branches are assumed to be vertical linear structures in the finger skeletons. 
The algorithm of the spanning tree based finger branch extraction is: 
\begin{description} 
\item [Step 1: ] We identify vertical structures from the Reeb graph of a given finger skeleton. From the graph, we search the longest downward path (i.e., the path with the longest height persistence) to be a new branch by using the breadth-first search. 
Next, we delete the points and edges of the new branch from the given graph, and repeat Step $1$ until the given graph becomes empty. After obtaining all the branches, we insert the longest one into a first-in-first-out (FIFO) queue, and mark this one to be enqueued. 

\item [Step 2: ] We build connections between branches. We obtain a branch from the queue, and identify which unmarked branches have edges connecting with it in the original graph; we then record such connecting edges. 
We insert the newly identified branches into the queue, and mark these branches to be enqueued. 
\end{description}
Fig.~\ref{fig:medium-finger}g shows the constructed branches and the connections between the branches. We visualize constructed branches by tree-based visualizations to minimize occlusion, such as Fig.~\ref{fig:simple-finger}d and Fig.~\ref{fig:medium-finger}h.

\subsubsection{Trim of Finger Skeletons with Removal of Short Branches and Cycles}
We trim the Reeb-graph based finger skeletons. When a finger structure is complex, essential geometric information may be occluded in the full skeleton. For example, in Fig.~\ref{fig:complex-finger}e, short branches occlude persistent branches and hinder the perception of the overall geometric structure. Hence, we prune the finger skeletons to preserve the most relevant geometric information of the fingers and minimize the occlusion (R2.3). The trim of the Reeb-graph based skeletons is based on the height persistence (R1.2). Suggesting by previous works discussed in Sect.~\ref{sect:reeb_graph}, we remove short branches and short loops to prune finger skeletons. 

The trimmed skeletons are shown in (f) of both Fig.~\ref{fig:medium-finger} and \ref{fig:complex-finger}. As a result of the trim, the geometric structure of fingers can be readily understood. 




\begin{figure}[htb]
\centering
 
  \includegraphics[width=0.80\columnwidth]{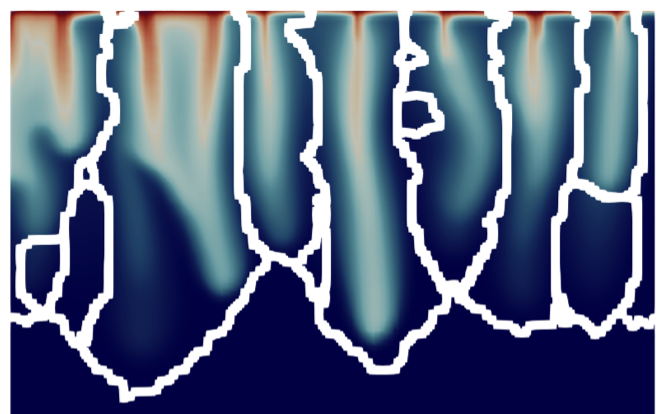}

  \caption{We show the segmented fingers, where the segmentation boundaries are represented by white lines.  }\label{fig:method-segmentation}
\end{figure}

\subsection{Extraction and Tracking of Finger Volume}
We extract complete finger volume, and track the volume of fingers and branches based on volume overlapping. 

\subsubsection{Volume Segmentation for Extraction of Complete Volume of Individual Fingers} \label{sect:segmentation}

We recover the complete volume of individual fingers by segmenting the 3D density field into connected subfields. We segment finger cores first. Almost all finger cores are connected through the high-density hexagonal cells in the top layer by observed from Fig.~\ref{fig:method-top-surface}, where the \textit{top layer} is the diffusive boundary layer contained in the top grid cells. Hence, when separating finger cores, we ignore the part of finger cores in the top layer, as suggested by the previous works \cite{aldrich2016viscous, lukasczyk2017viscous, luciani2018details}. We select a fixed height value as the separation between the top layer and the bottom layer of the whole domain through visual inspection by following the previous works \cite{aldrich2016viscous, lukasczyk2017viscous, luciani2018details}. Given the height of the whole domain ranging from $40$ to $0$ in this application, we find that the fingers are separated well throughout all timesteps for a fixed height value of $38$. For other datasets, our visual-analytics system provides visualizations for scientists to identify a suitable value. We then obtain each individual finger core as a connected component of the finger core voxels, excluding the top layer. After that, we use the watershed traversal method \cite{vincent1991watersheds} to extract the complete volume of individual fingers from the 3D regions with non-zero densities by expanding from individual finger cores. 



We display boundaries between the segmented fingers in Fig.~\ref{fig:method-segmentation} by using white lines. Fig.~\ref{fig:simple-finger}e presents the extracted voxels of a finger; more complex examples are displayed in (c) and (d) of both Fig.~\ref{fig:medium-finger} and Fig.~\ref{fig:complex-finger}. The segmented fingers satisfy the cognitive requirement of the earth scientist regarding individual fingers.

\subsubsection{Volume Overlapping Based Tracking of Fingers and Branches} \label{method:tracking}

We track fingers and their branches (R3). The volume overlapping based tracking method \cite{silver1996volume} is adopted because fingers usually flow downward and do not move with a significant horizontal deviation. We first relate fingers between consecutive timesteps with volume overlapping (i.e., that share grid cells). The number of shared cells and densities of these shared cells reflect the strength of connections between related fingers. Also, the shared cells have positions that reveal where the volume of the fingers overlaps. After identifying the correspondence between fingers, we further relate branches of corresponding fingers also based on the volume overlapping.

\section{Spatio-Temporal Visualizations of Viscous and Gravitational Fingers}
\begin{figure*}[hbt]
 \centering 
 \includegraphics[width=\linewidth]{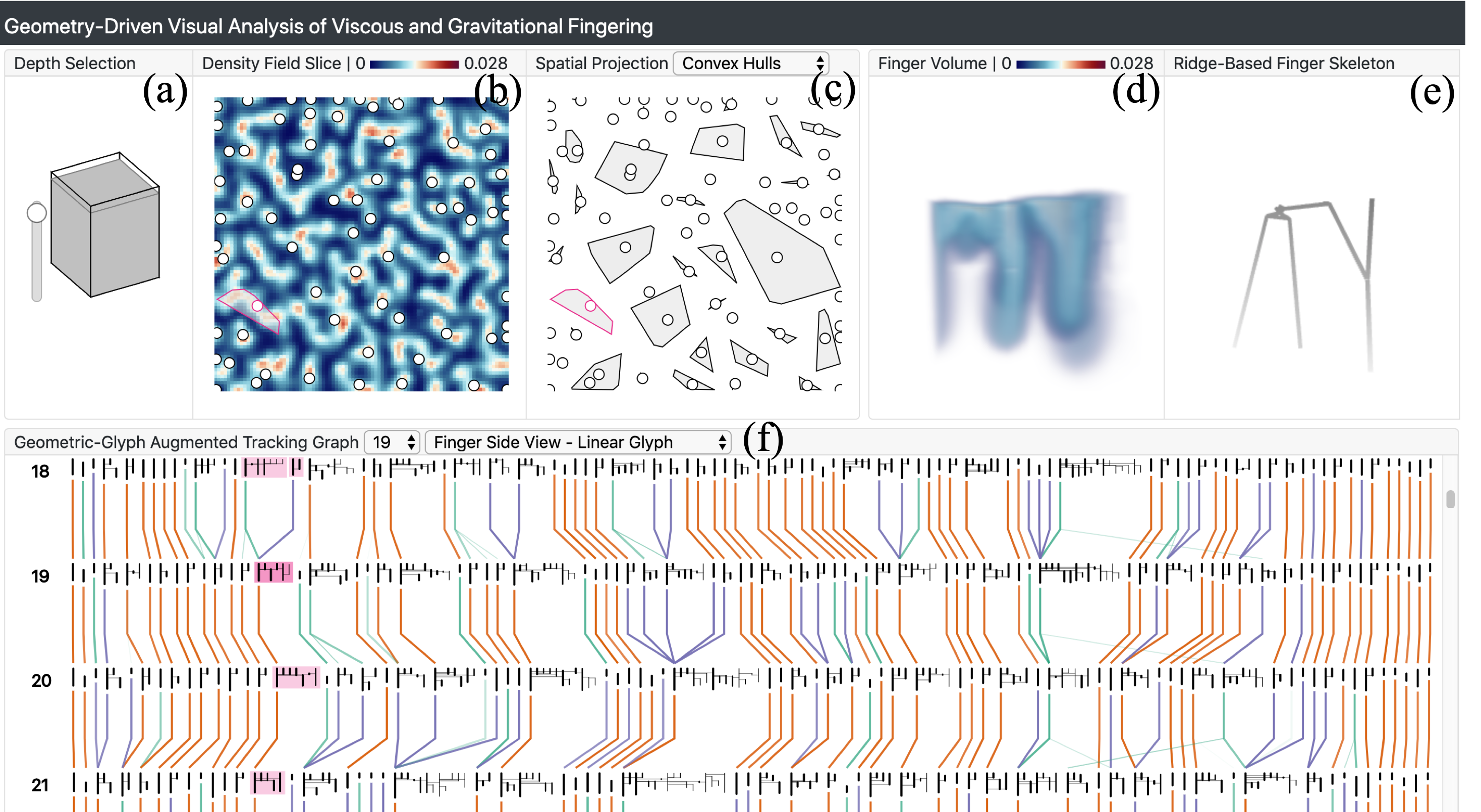}
 \caption{A geometry-driven visual-analytics system is used to study viscous and gravitational flow instabilities. (a) Depth slider is used to specify a depth of interest. The 3D domain under the specified depth is shown for analysis. (b) View of a density slice at the specified depth, which also shows centroid points of fingers projected onto the $x$-$y$ plane. (c) The spatial projection view displays the fingers projected onto the $x$-$y$ plane using scattered convex hulls. (d) The finger volume view displays the density field of a selected finger. (e) The finger skeleton view shows the geometric structure of a selected finger in 3D space. (f) Geometric-glyph augmented tracking graph visualizes the geometric evolution of fingers based on geometry-driven planar glyphs. 
 }
  \label{fig:visual-analysis-system}
\end{figure*}

We create an interactive visual-analytics system to perform spatio-temporal analyses on the geometric structures of the viscous and gravitational fingers. In our system, we present juxtaposed visualizations so that users can compare fingers over space and time. Fig.~\ref{fig:visual-analysis-system} shows the complete visual-analytics system, which consists of six panels marked as (a)-(f). Regarding spatial exploration (R2), (a) the depth selection panel allows users to select a depth of interest in the spatial domain.  Then, (b) the density field slice view displays the density field at a selected depth as a 2D heatmap. (c) the spatial projection panel presents a high-level spatial view, and uses convex hulls to indicate the extent of fingers projected on the $x$-$y$ plane.  Users can also observe different geometric features (R1) of fingers in detail through (d) a volume rendering image and (e) a 3D skeleton visualization. Finally, at the bottom of Fig.~\ref{fig:visual-analysis-system}, (f) the geometric-glyph augmented tracking graph displays the evolution of fingers (R3). All the views in our system are interactive and coordinated together to enable coherent visual analyses. In the following, we describe each of these panels in detail.




\begin{figure}[tb]
  \centering 
  \includegraphics[width=\columnwidth]{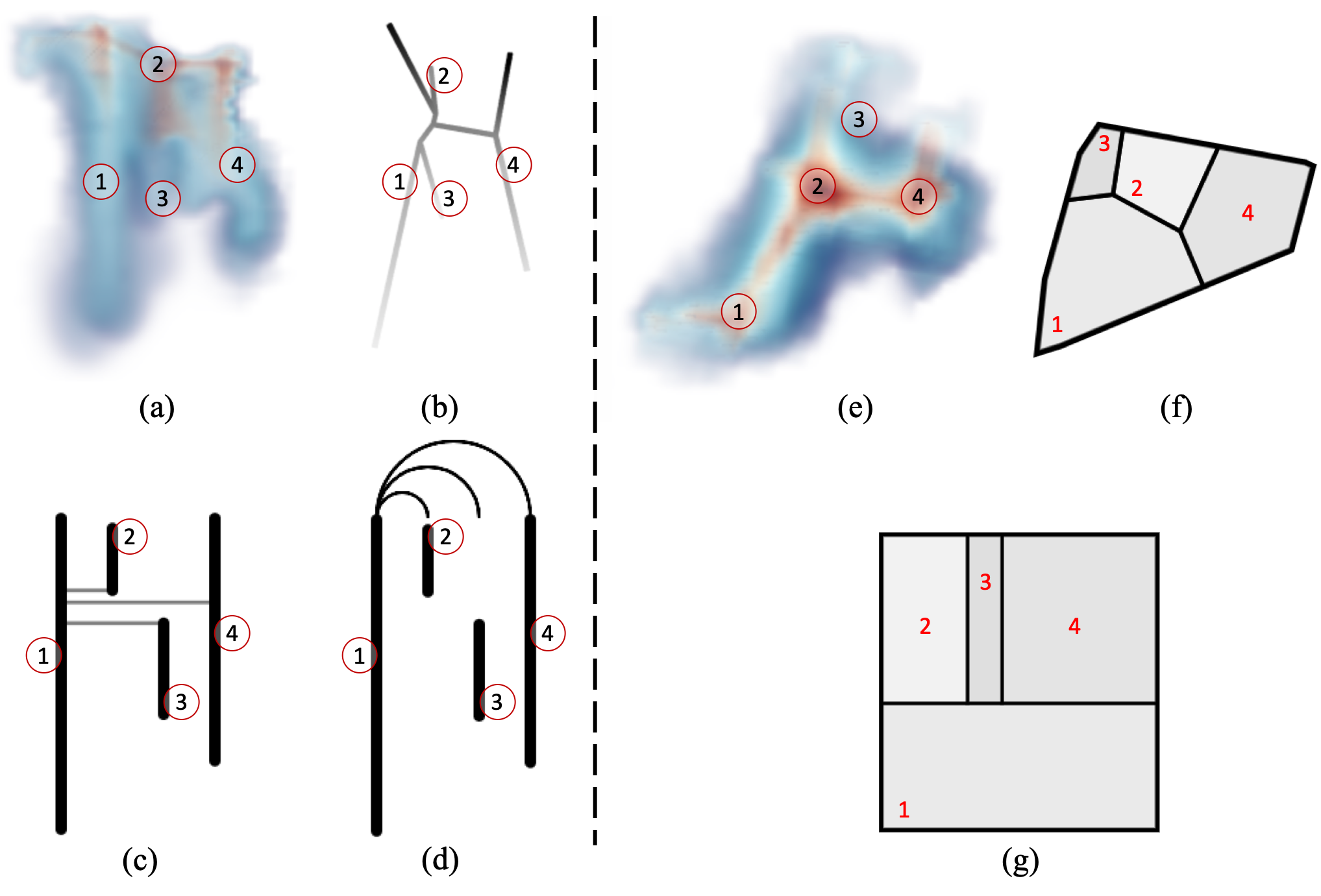}
  \caption{We showcase glyph designs of a finger. The same numbers label corresponding branches. The left four views display side views of the finger. Image (a) shows the side view of the finger volume. Image (b) is the skeleton of the finger. The linear glyph (c) displays connections between branches of the finger. When horizontal connections are too close or even overlap, the alternative design (d) shows the tree structure with less ambiguity than (c). Right three views display top views of the finger. Image (e) is the top view of the finger volume. The Voronoi glyph (f) is shown on the spatial projection panel; Voronoi cells that are inside the convex hull are corresponding to the finger branches. The rectangular glyph (g) is for contrasting the topological complexity between the finger branches on the tracking graphs. 
  }
  \label{fig:designs-showcase}
\end{figure}


\subsection{Spatial Visualizations of Fingers}

\subsubsection{Depth Selection and Density Field Slicing}
Density field slicing is an important tool for scientists to study the change of densities inside fingers in space (R2). In the depth selection panel (Fig.~\ref{fig:visual-analysis-system}a), our system allows users to drag a slider to select a depth of interest; the 3D spatial region below the selected depth is highlighted by gray color in a 3D cube. The density field slice at the selected depth is then extracted and shown as a 2D heatmap in Fig.~\ref{fig:visual-analysis-system}b. The finger volume and skeleton views (Fig.~\ref{fig:visual-analysis-system} d and e) are updated to show finger structures under the selected depth only.  The selected depth is set to the top of the 3D domain initially, because \ce{CO2} is injected from the top of the aquifer. 


\subsubsection{Spatial Projection of Fingers} \label{vis:projection}

A high-level spatial projection view is necessary to reveal how fingers distribute horizontally (R2.2). Convex hull and Voronoi treemap are further applied to display the projected individual fingers, as detailed below. 


\textbf{Convex hull: } We project critical points of finger skeletons onto the $x$-$y$ plane. We then draw a convex hull to enclose the projection of the critical points of each finger. The convex hull representation, shown in Fig.~\ref{fig:visual-analysis-system}c, demonstrates the coverage of each finger in space, and help scientists identify large fingers instantly. This view also supports the selection of a finger of interest to display the volume and skeleton of the selected finger. The convex hull of a selected finger is also superimposed on the density field slice view, as shown in the lower-left corner of Fig.~\ref{fig:visual-analysis-system}b to confirm the existence of the selected finger in the density field. In addition, we project centroid points of finger skeletons onto the $x$-$y$ plane for both the density field slice and spatial projection view. 

\textbf{Voronoi treemap: } To reveal information of finger branches (R1.1), we use the Voronoi treemap \cite{balzer2005voronoi} to represent each finger branch by a Voronoi cell, and embed the Voronoi cells in the convex hulls of fingers instead of centroid points as shown in Fig.~\ref{fig:designs-showcase}f. 
The relative positions of branches of a finger are revealed by the positions of Voronoi cells within the convex hull of the finger. The extent of a branch is represented by the cell extent. 
Dark Voronoi cells mean that the corresponding finger branches exist in the deep region of the 3D domain. More precisely, the darkness of Voronoi cells encodes the average depth of critical points that are in the corresponding finger branches. 
Users can select to observe either centroid points for simplicity or Voronoi cells for details of finger branches.


\subsubsection{3D Spatial Visualizations of a Selected Finger} \label{sect:spatial_visualization_selected_finger}

We provide 3D volume- and skeleton-based visualizations for geometric analyses (R1) and spatial exploration (R2) after selecting a finger of interest. Users can interactively rotate both of the volume and the skeleton to study a selected finger from different viewpoints, and zoom into the representations to study details of the finger structure. 

\textbf{Volume visualization: }
The volume rendering image of an individual finger, as shown in Fig.~\ref{fig:visual-analysis-system}d, visualizes how the density of the finger distributes in the physical domain, which allows scientists to interpret the finger intuitively. Moreover, the ray casting based volume rendering remedies the occlusion (R2.3) on the direct display of voxels (e.g., Fig.~\ref{fig:complex-finger}c) with transparency. 





\textbf{Geometric skeleton visualization: }
To visualize the overall geometric structures of fingers in 3D space (R1) and analyze how fingers grow vertically (R2.1), we display finger skeletons on the screen using orthogonal projection such as Fig.~\ref{fig:designs-showcase}b.  To recover the density information of fingers, we use intensity gradients along lines to indicate the density changes of finger branches. As a result, high-density branches are highlighted, and low-density branches are under-emphasized. 

\subsection{Geometric-Glyph Augmented Tracking Graph for Temporal Visualization of Fingers} \label{sect:tracking_graph}

\begin{figure*}[hbt]
  \centering 
  \includegraphics[width=\linewidth]{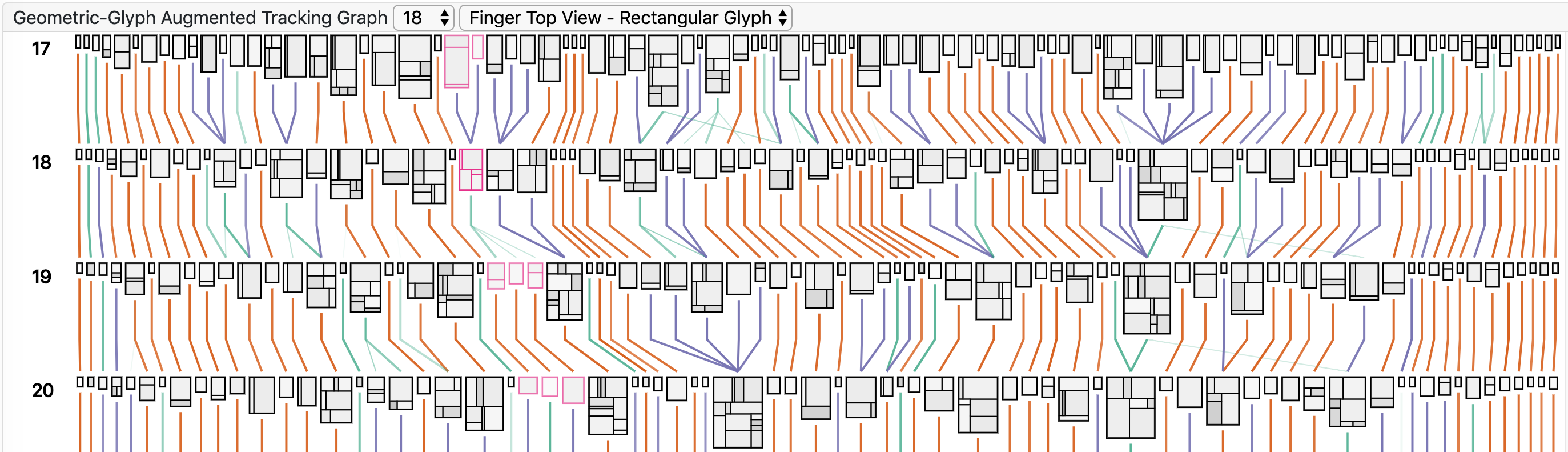}
  \caption{We display the geometric-glyph augmented tracking graph with rectangular glyphs. A finger associated with merging and splitting events at timestep $18$ is highlighted by red-violet and analyzed in Fig.~\ref{fig:merge-split} further. }
  \label{fig:tracking-graph-treemap}
\end{figure*}

The geometric-glyph augmented tracking graphs, as shown in Fig.~\ref{fig:visual-analysis-system}f and Fig.~\ref{fig:tracking-graph-treemap}, visualize the evolution of fingers to facilitate temporal exploration of fingers (R3). The tracking graphs help scientists identify various evolutionary events of fingers, and analyze how the evolutionary events happen in detail through interactions (R3.1). Each row of the tracking graph displays the fingers at one timestep; the timestamp is labeled at the left of the panel. Widths of finger glyphs are adjusted according to the topological complexity of finger structures; more complex fingers have wider space for drawing. Users can choose to filter out the glyphs of short fingers if tracking persistent fingers is preferred. Also, users can switch the style of the glyph between the linear glyph (Fig.~\ref{fig:designs-showcase}c) and the rectangular glyph (Fig.~\ref{fig:designs-showcase}g) to observe different facets of fingers. 
Below we describe the tracking graph in more detail, including the generation of glyphs for showing the geometric structures of fingers, the color of links encoding the evolutionary events of fingers, the minimization of link crossings for reduction of visual clutter, and the interactions for exploration of temporal events. 

\subsubsection{Linear Glyph for Finger Side View} \label{vis:linear_glyph}

To display the evolution of geometric structures of fingers (R3.1), we nest geometry-driven glyphs on the tracking graphs. To visualize the side view of fingers over time, we draw finger branches and their connections in a plane (R1.1) with minimized occlusion (R2.3) and also preserve the height of branches (R1.2). Two possible designs are discussed in the following. 

\textbf{Projection: } We may project finger skeletons onto a plane. However, the traditional orthogonal projection suffers from edge crossing problems.  To remedy the edge crossing for the projection of tree-like structures, Marino and Kaufman \cite{marino2016planar} produced radial planar embeddings of the structures. However, fingers are vertical objects, and radial planar embeddings of fingers may lose the depth of fingers. Although the method of Marino and Kaufman \cite{marino2016planar} is good at preserving the curvature of 3D objects, the curvature is not an essential feature of fingers. 

\textbf{Tree drawing: } We draw fingers as linear glyphs to capture abstract forms of fingers, as an example illustrated in Fig.~\ref{fig:designs-showcase}c. Heine et al. \cite{heine2011drawing} proposed a graph-drawing method to plot branches of contour trees in a plane with minimized edge crossings (R2.3). The method of Heine et al. \cite{heine2011drawing} represents branches by vertical lines and denotes links between branches by horizontal lines. Connections between branches (R1.1) and vertical features of branches (R1.2 and R2.1) are shown clearly in the results of this method. Thus, we augment the tree-drawing method of Heine et al. \cite{heine2011drawing} to display the side view of fingers, such as in Fig.~\ref{fig:designs-showcase}c. Note that, although trees are planar graphs, crossings of tree edges are inevitable for some instances by using this design; the reason is that only the $x$-axis is free to arrange tree branches, and the $y$-axis is used to encode the height of branches. We draw the longest branch of a finger at the leftmost of the glyph so that we can locate the principal branch quickly. Users can choose to encode change of densities along branches by using the gradient darkness of vertical lines such as in Fig.~\ref{fig:simple-finger}d. When hovering over a linear glyph on the tracking graph, connections between branches of this finger become arcs hanging at the top of the glyph to reduce visual clutter (R2.3), such as in Fig.~\ref{fig:designs-showcase}d. The corresponding tracking graph is shown in Fig.~\ref{fig:visual-analysis-system}f. 








\subsubsection{Rectangular Glyph for Finger Top View} \label{vis:treemap}
We draw the top view of fingers to display quantitative geometric attributes (R1) and relative positions (R2.2) of finger branches by using treemap based rectangular glyphs. An example is illustrated in Fig.~\ref{fig:designs-showcase}g. The treemap technique \cite{johnson1991tree} maps elements onto a rectangular region in a space-filling manner and displayed quantity values of elements effectively. The spatially ordered treemap \cite{wood2008spatially} extended from squarified treemaps \cite{bruls2000squarified} arranges the rectangles of elements based on both the spatial proximity and the balance of aspect ratio, and, can display the spatial distribution of elements (R2.2). Hence, we use the spatially ordered treemap \cite{wood2008spatially} to generate rectangular glyphs for finger branches, as shown in Fig.~\ref{fig:designs-showcase}g. Each branch is represented by a rectangle on the glyphs. Areas of rectangles are able to encode the numeric attributes of branches, including statistics or topological measurements. In this work, the area encodes the topological complexity of the corresponding finger branch (R1.1). Intuitively, a finger branch having more critical points usually indicates that this finger branch is connected to more other branches and thus is more complex in terms of its topological structure. Hence, we define \textit{topological complexity of a finger branch} by the number of critical points on the branch skeleton. 
Moreover, to compare fingers in the tracking graph and assign more space for topologically complex fingers, we encode the size of rectangles of whole fingers by the topological complexity of fingers. The \textit{topological complexity of a finger} is defined by the number of critical points on the finger skeleton; intuitively, a finger has more critical points, usually indicates this finger has more branches and hence is more complex in its topological structure. The corresponding tracking graph is shown in Fig.~\ref{fig:tracking-graph-treemap}.

\subsubsection{Link Encoding for Evolutionary Events of Fingers} \label{vis:link_encoding}

Links represent the temporal relationships between fingers. 
When time varies, fingers may grow, merge with other fingers, or split into multiple fingers. To indicate these three types of finger evolution, we use three hues of links following the qualitative color advice of ColorBrewer \cite{brewer2018colorbrewer}.
\begin{description} 
\item [Brown link: ] Brown links indicate that fingers grow with minor changes between consecutive timesteps. Specifically, at least seventy-five percent volume of the finger is preserved in one of the connected fingers at the subsequent timestep. 

\item [Blue link: ] A blue link indicates the case when multiple fingers merge into one finger at the subsequent timesteps. Specifically, the finger at the subsequent timestep incorporates seventy-five percent volume of each of the fingers at the previous timestep. 

\item [Green link: ] A green link indicates that a finger splits into multiple fingers at the subsequent timestep, and none of the fingers at the subsequent timestep have seventy-five percent volume of the finger at the previous timestep. 
\end{description}
We use the opacity of links to encode link weights. The \textit{weight of a link} is the size of the overlapping volume between linked fingers. If fingers have weak connections, their links are almost transparent so that visual clutter is reduced.

\subsubsection{Iterative Minimization of Link Crossings} \label{vis:link_crossing}

We reduce link crossings to alleviate the visual clutter of the tracking graphs. Since links are weighted, the reduction of link crossings between every two rows of glyphs is a graph drawing problem: edge crossing minimization for weighted bipartite graphs.  \c{C}ak{\i}ro$\bar{g}$lu et al. \cite{ccakiroḡlu2007crossing} enhanced and tested five heuristic methods for this graph drawing problem, and concluded that W-GRE \cite{yamaguchi1999approximation, ccakiroḡlu2007crossing}
method produces the best results. Hence, we utilize W-GRE method in our application to minimize intersections of weighted links. 

We propose a W-GRE based iterative algorithm for the crossing minimization of multiple rows, which arranges finger glyphs of each row iteratively until obtaining a good result. We describe the algorithm in the following. 
\begin{description} 
\item [Step 1: ]The finger glyphs in the first row are arranged by the ascending order of the $x$ coordinates of the finger centroids initially. 
\item [Step 2: ]We order glyphs from the second row to the last row. We minimize the crossings of the links between a given row and its previous row by using the W-GRE \cite{ccakiroḡlu2007crossing} method. 
\item [Step 3: ]We order glyphs from the second-to-last row to the first row. We minimize the crossings of the links between a given row and its following row by using the W-GRE \cite{ccakiroḡlu2007crossing} method. 
\item [Step 4: ]Repeat Step 2 and Step 3 to order the fingers of each row until reaching the convergence of the finger order. Usually, repeating two times can produce a good result in our experiments. 
\end{description}

\subsubsection{Interactive Finger Tracking} \label{vis:interaction}

We offer three interactions for scientists to track fingers. 

\textbf{Finger selection: } Scientists can select a finger of interest by click on the finger glyph from the tracking graphs. Then, the glyphs of the selected finger and the other related fingers are highlighted in the tracking graphs by coloring the backgrounds (Fig.~\ref{fig:visual-analysis-system}f) or the strokes (Fig.~\ref{fig:tracking-graph-treemap}). Additionally, the details of the selected finger are displayed in the volume and skeleton views. 


\textbf{Finger volume tracking: } The earth scientist is also interested in tracking the volume of a finger. Note that each link is associated with the overlapping volume between temporally related fingers. Users can click on a finger of interest first and click on one of the associated links to observe overlapping volume between this finger and the other finger in the finger volume view, which is illustrated in Sect.~\ref{sect:finger_evolution} and Fig.~\ref{fig:merge-split} in detail. 

\textbf{Branch tracking: } The earth scientist requires designs to track geometric structures of fingers (R3.1). If a complex finger separates into multiple smaller fingers over time, it is essential to track each branch of the complex finger comes to which smaller finger entity afterwards; also, if multiple fingers fuse into a complex finger, it is important to identify the correspondence between the individual fingers and the branches that they got merged to. On our tracking graph, we identify corresponding branches between linked fingers interactively.
When hovering over a link between two connected fingers, the relevant branches between the fingers are highlighted by the red-violet color on glyphs, as shown in Fig.~\ref{fig:merge-split}. 










\section{Case Studies and Scientist Feedback} \label{sect:case_studies}

As we were developing our system, we were in close contact with \remark{two earth scientists including the one who is referred to in Sect.~\ref{sect:expert_requirements}}. In the following, we discuss the use cases that were studied by the earth scientists when they used our geometry-driven visual-analytics system to explore the spatio-temporal phenomena of viscous and gravitational fingers. Also, we present their feedback and suggestions after they performed domain-specific tasks.




\begin{figure}[htb]
  \centering 
  \includegraphics[width=\columnwidth]{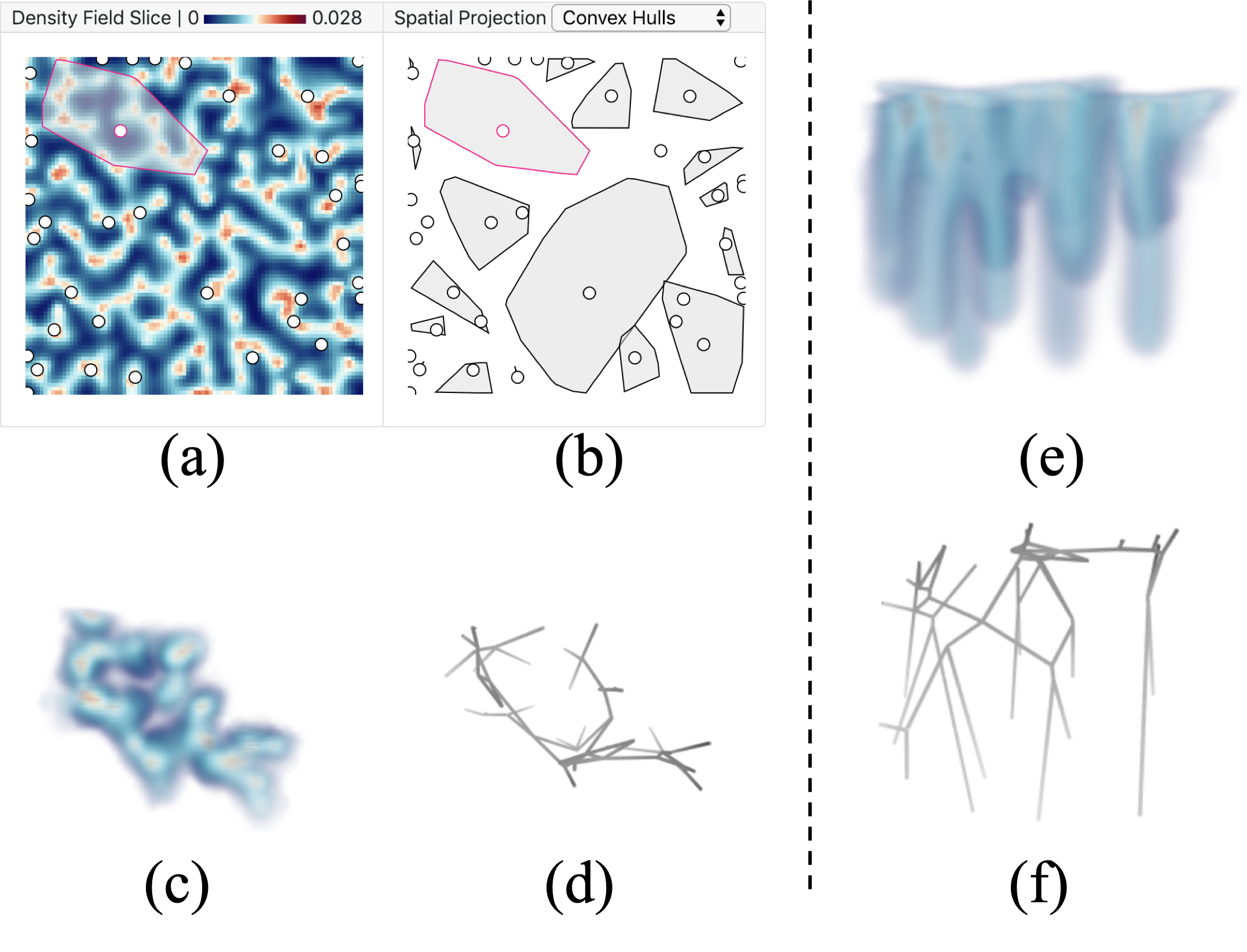}
  \caption{The earth scientists identified a finger of interest from (b) the spatial projection panel. Then, they observed the high-density hexagons under the convex hull of the finger in (a) the density field slice view. They further confirmed the correspondence between (c) the top view of the finger volume and the hexagons in (a). Also, they found the finger skeleton (d) preserved the geometric structure of (c). Next, they looked at the side view of the finger volume (e) but felt difficult for observing finger branches due to the occlusion. They then interacted with the 3D skeleton (f) and obtained how these branches form in space over time. 
  }
  \label{fig:complex-finger-case}
\end{figure}

\subsection{Case 1: Spatial Analysis of Geometric Features of Fingers}
Because the existing methods have limitations in capturing geometric patterns of fingers, the earth scientists mentioned the difficulty in identifying branching structures of fingers in 3D space. 
There are two challenging problems. The first is to identify where the fingers and branches are in space. After the identification, the second problem is to comprehend how the high-density hexagonal sheets in the top (e.g., Fig.~\ref{fig:complex-finger-case} c and d) split and develop into columnar fingers (e.g., Fig.~\ref{fig:complex-finger-case} e and f) spatio-temporally. Specifically, given hexagonal cells (e.g., Fig.~\ref{fig:complex-finger-case} c and d), it is important to know that whether the fingers tend to form along with the boundaries of hexagonal cells or form below the centers of the hexagonal cells. By using our system, the earth scientists were equipped to solve the problems efficiently. 


The earth scientist identified where are the fingers and branches by observing spatial visualizations. 
They first looked at the density field slice view, Fig.~\ref{fig:complex-finger-case}a. The scientists thought the projection points added on the view is an effective addition to the traditional density-slice views, which allowed them to identify finger locations on any slice with the corresponding hexagonal features at the slice. Furthermore, from the spatial projection panel, Fig.~\ref{fig:complex-finger-case}b, the earth scientists found two fingers that occupy a large space. They clicked on one of the two for analysis, and the convex hull of the finger was superimposed on the density field slice view, Fig.~\ref{fig:complex-finger-case}a. They focused on the high-density hexagons under the convex hull in the slice view, and adjusted the $z$-value slider of 
to understand the change of densities along the $z$ coordinate. Note that the finger volume (Fig.~\ref{fig:complex-finger-case}c) and skeleton (Fig.~\ref{fig:complex-finger-case}d) views were displayed after the selection of the finger. They verified the correspondence between the hexagons under the convex hull in Fig.~\ref{fig:complex-finger-case}a and the top view of the volume, Fig.~\ref{fig:complex-finger-case}c. Next, they confirmed the correspondence between the finger volume and skeleton. The skeleton Fig.~\ref{fig:complex-finger-case}d captured the geometric structure of the volume Fig.~\ref{fig:complex-finger-case}c. Afterwards, they rotated the finger volume and skeleton to see side views of the finger. The finger branches occluded severely in the volume visualization Fig.~\ref{fig:complex-finger-case}e. However, when interacting with the skeleton in Fig.~\ref{fig:complex-finger-case}f, they intuitively acquired the finger branches in space. Hence, they thought the extracted finger skeletons (e.g., Fig.~\ref{fig:complex-finger-case}f) were effective in identifying the branching structures of fingers. 

The scientists afterwards understood how \ce{CO2} volume flows and how fingers form in space by using our system. Since fingers result from \ce{CO2} flows in space over time, by observing the skeleton (Fig.~\ref{fig:complex-finger-case}f), the earth scientists quickly captured the downward flowing track of \ce{CO2} by following the skeleton and comprehended how the finger and its branches grow. Moreover, from the finger skeletons (e.g., Fig.~\ref{fig:complex-finger-case}f), the experts obtained an insight that the fingers usually formed along the boundaries of hexagonal cells rather than forming below the centers of hexagonal cells. 


\begin{figure}[htb]
\centering
\captionsetup[subfigure]{labelformat=empty}

 \begin{subfigure}[b]{0.3\columnwidth}
   \includegraphics[width=\columnwidth]{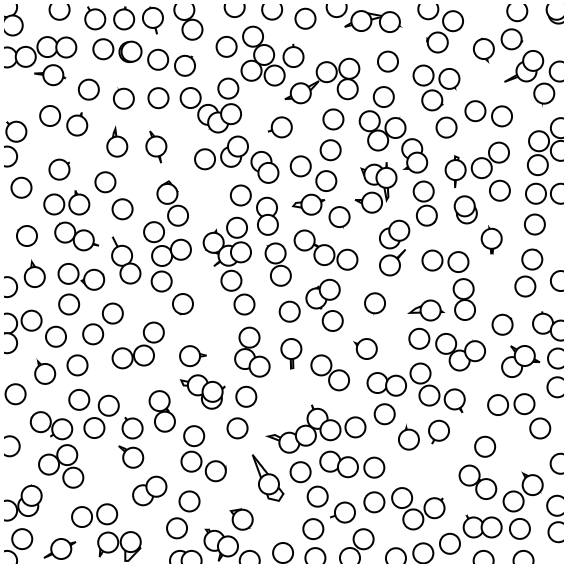}
   \caption{(a$_1$)}
 \end{subfigure}
 \hfill
 \begin{subfigure}[b]{0.3\columnwidth}
   \includegraphics[width=\columnwidth]{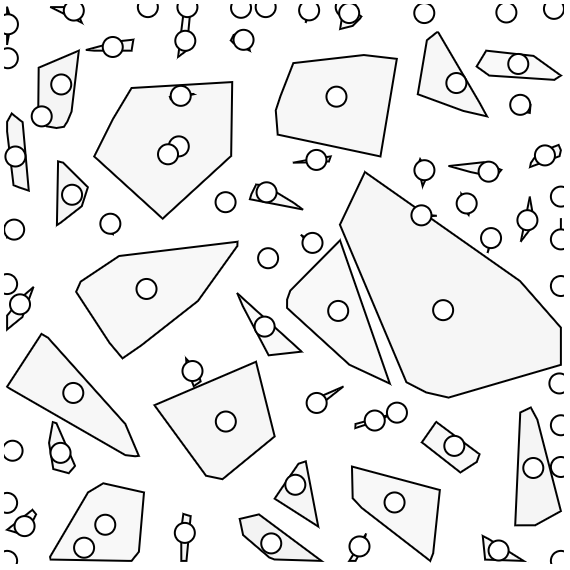}
   \caption{(b$_1$)}
 \end{subfigure} 
 \hfill
 \begin{subfigure}[b]{0.3\columnwidth}
   \includegraphics[width=\columnwidth]{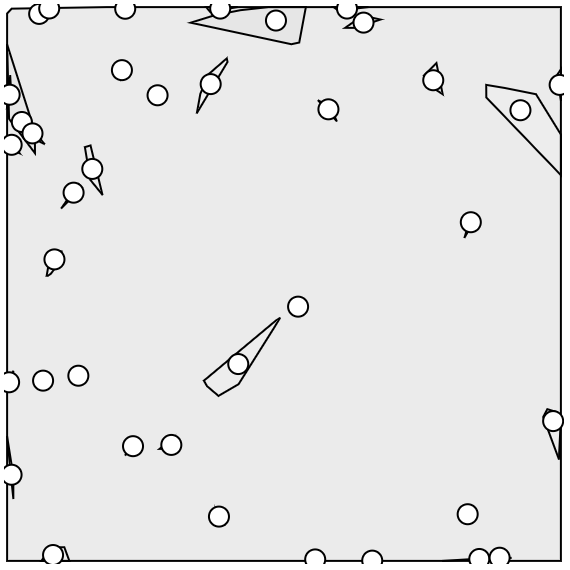}
   \caption{(c$_1$)}
 \end{subfigure} 

 \bigskip

 \begin{subfigure}[b]{0.3\columnwidth}
   \includegraphics[width=\columnwidth]{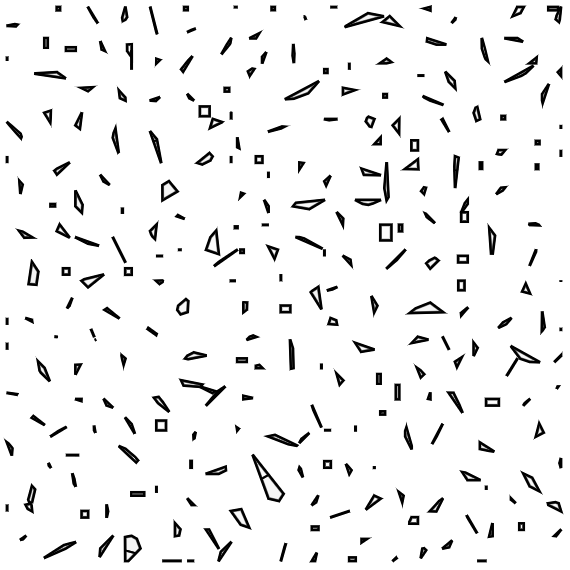}
   \caption{(a$_2$)}
 \end{subfigure}
 \hfill
 \begin{subfigure}[b]{0.3\columnwidth}
   \includegraphics[width=\columnwidth]{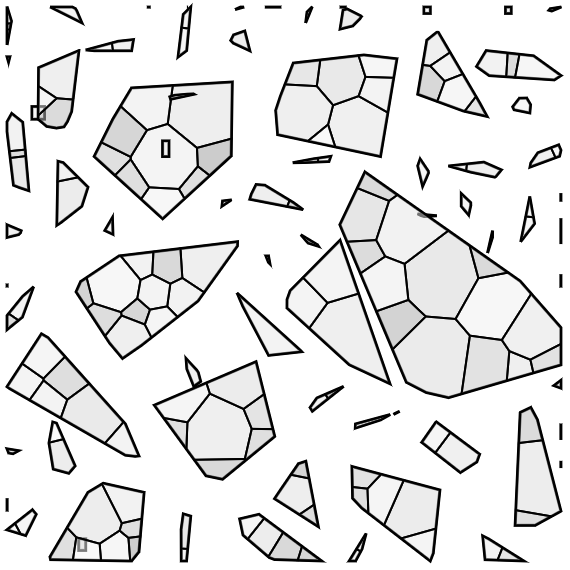}
   \caption{(b$_2$)}
 \end{subfigure} 
 \hfill
 \begin{subfigure}[b]{0.3\columnwidth}
   \includegraphics[width=\columnwidth]{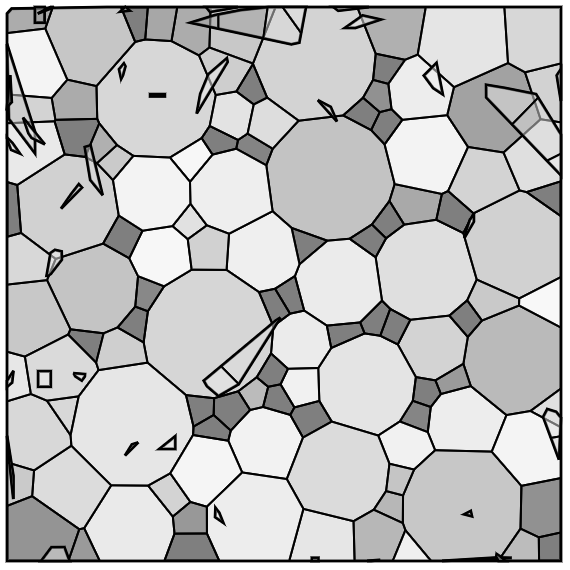}
   \caption{(c$_2$)}
 \end{subfigure} 

  \caption{There are three important phases for the evolution of fingers: (a) the onset phase, (b) the growth phase, and (c) the termination phase. Convex hulls of (a$_1$), (b$_1$), and (c$_1$) display the spatial coverage of individual fingers. Voronoi glyphs of (a$_2$), (b$_2$), and (c$_2$) show finger branches. }

  \label{fig:phases}
\end{figure}



\subsection{Case 2: Temporal Analysis of Fingers}
We present how to perform temporal analysis of fingers by using the geometry-driven visual-analytics system. 

\subsubsection{Case 2.1: Recognition of Evolutionary Phases}
According to the earth scientists, the evolution of the instability has (at least) three important phases: onset, growth, and termination. 
The earth scientists identified the three phases based on the spatial projection panel, as shown in Fig.~\ref{fig:phases}. In the onset phase (Fig.~\ref{fig:phases} a$_1$ and a$_2$), the instability is triggered, and the displacement front breaks up into many small-scale fingers. After the onset phase, the dense fingers grow non-linearly and penetrate the underlying lighter fluid. In Fig.~\ref{fig:phases} b$_1$ and b$_2$, many medium-sized fingers form in this growth phase. These medium-sized fingers grow, merge, and/or split over time. Eventually, multiple fingers merge to form a few `mega-plumes' that span most of the reservoir, as shown in Fig.~\ref{fig:phases} c$_1$ and c$_2$. Once those fingers reach the bottom of the domain, the instability shuts down or terminates.



\begin{figure}[htb]
  \centering 
  \includegraphics[width=\columnwidth]{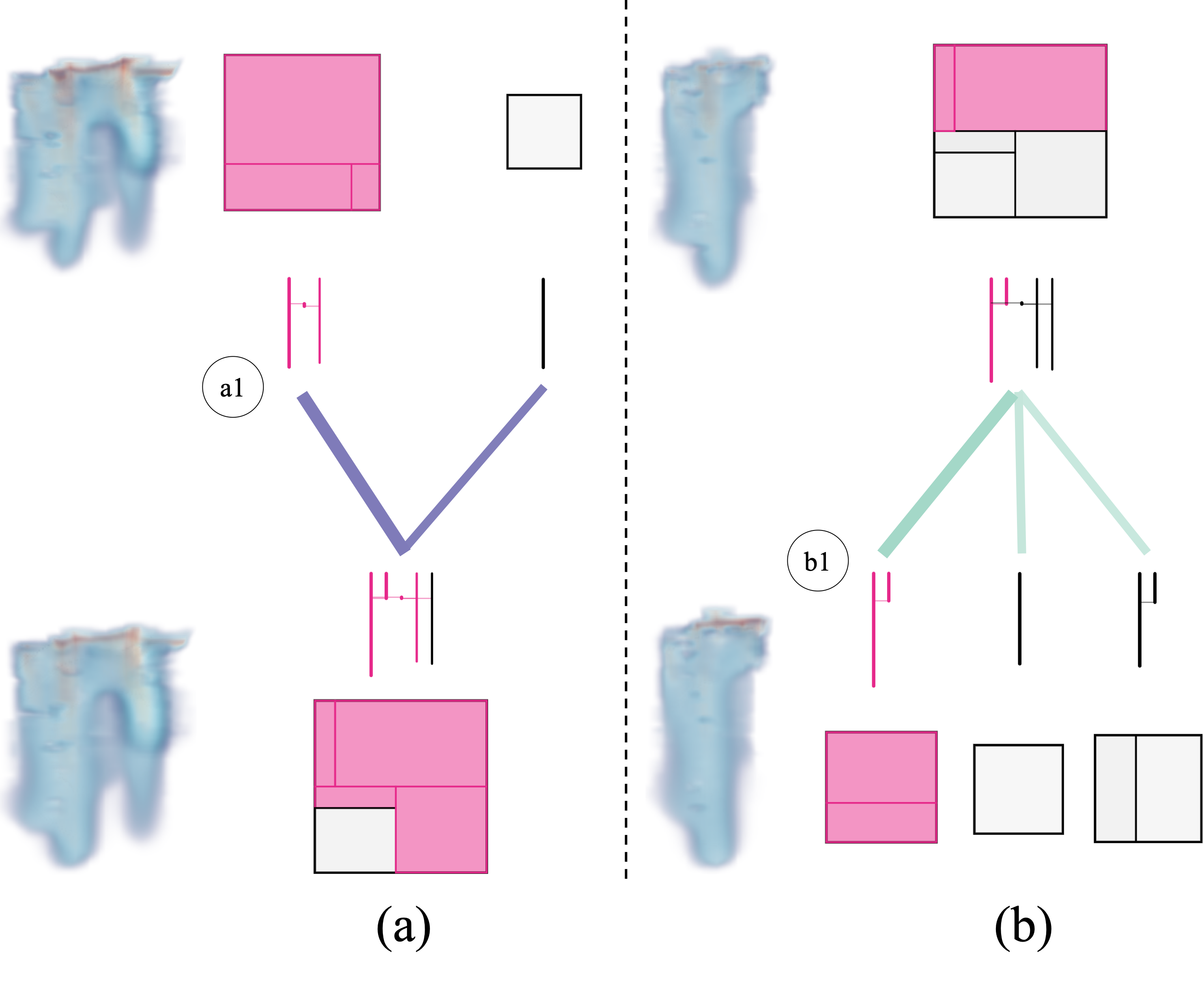}
  \caption{After evolving one timestep, (a) fingers may merge into a complex finger, or (b) a complex finger may split into more fingers. To track branches, users can hover over a link; the link becomes thick, and associated branches on glyphs are highlighted by red-violet color. To track the volume of fingers, users can click on a link; then, the overlapping volume between linked fingers is shown to be compared. 
  }
  \label{fig:merge-split}
\end{figure}


\subsubsection{Case 2.2: Exploration of Evolutionary Events} \label{sect:finger_evolution}
The earth scientists identified the growth, merging, and splitting of time-varying fingers from the geometric-glyph augmented tracking graph. During our evaluation session, first, we illustrated the encodings of our designs to the earth scientists. 
After we explained the three colors of links used to represent the growing, merging, and splitting events, they appreciated the temporal evolution of the fingers was intuitively displayed.  
We next explained the two kinds of finger glyphs to the earth scientists. Compared with previous tracking graphs \cite{aldrich2016viscous, lukasczyk2017viscous, luciani2018details} that only use dots to represent fingers, the geometric glyphs of our tracking graphs offered more details and facets of fingers to the earth scientists. With respect to the linear glyphs (e.g., Fig.~\ref{fig:designs-showcase}c), at first glance, they were confused with the encoding of horizontal lines initially. However, after our explanation, they were able to understand the horizontal lines representing connections between vertical branches and thought this design was effective. 
They thought drawing fingers in a plane would cause distortion, but found that the branches and their connections were shown clearly, and it was easy to count the number of vertical branches. In regards to the rectangular glyphs (e.g., Fig.~\ref{fig:designs-showcase}g), the expert captured the horizontal distribution of branches and quickly counted the number of complex branches. Also, they suggested highlighting the correspondence between the rectangles and the branches in the volume so that they could understand how the structures of fingers from the glyphs better. 

Here we illustrate how the earth scientists analyzed the evolution of fingers and branches by using our system. First, they found a finger of interest with merging and splitting events, which is highlighted by red-violet in Fig.~\ref{fig:tracking-graph-treemap} and extracted in Fig.~\ref{fig:merge-split}. Concentrating on the merging event of this finger, they hovered on the left link above the merged finger (Fig.~\ref{fig:merge-split}a), and hovered on the leftmost link below the finger (Fig.~\ref{fig:merge-split}b) to track the corresponding branches that were highlighted by red-violet in glyphs. The corresponding linear glyphs and rectangular glyphs between consecutive timesteps were found to be consistent. To track the volume, they first clicked on (a1), the upper-left finger of Fig.~\ref{fig:merge-split}a, to look at its volume; then, they clicked on the merged finger at the next timestep and clicked on the link to (a1) to observe the partial volume of the merged finger that comes from (a1). The two volume images are shown on the left of Fig.~\ref{fig:merge-split}a, and the temporal change of the corresponding volume was observed. Similarly, they tracked the overlapping volume of fingers in the splitting event, Fig.~\ref{fig:merge-split}b, and observed that the corresponding branches grew longer. After using our system, the earth scientists thought our tracking graphs were valuable to identify when and where the fingers grow, merge, and split over time. Compared with the previous works that do not construct branches of fingers explicitly, our methods tracked finger branches more efficiently.


\subsection{Scientist Feedback}
\remark{After using our system, the earth scientists thought that our geometry-driven system provided a new perspective on the analysis of viscous and gravitational fingering. }
Specifically, the visual-analytics system helped scientists acquire branches in space and branching behaviors over time efficiently and effectively, which relieved cumbersome work for the geometric analyses of fingers manually, since the branch tracking offered by our system is new and not directly available by using previous methods \cite{aldrich2016viscous, favelier2016visualizing, lukasczyk2017viscous, luciani2018details}. They also acknowledged that the extraction of fingers by our method was effective and that the minimal occlusion in visualizations leading to clear representations of fingers. 

\section{Discussion: Limitation and Future Work}

\textbf{Sensitivity of the ridge voxel detection: } 
If the quality of the data is low, for example, resulting from noise or coarse discretization, the approximation of derivatives may have higher errors leading to inaccurate detection of the ridge voxels. In addition, if the density fields are not smooth enough, the detected ridge structures can be fragmented. Although the quality of the data used in this paper is sufficient to extract high-quality ridges, methods for improving the data quality like noise removal could be used to improve the stability of the ridge detection algorithm. 
\remark{Moreover, as detected ridges are disconnected for other applications, one may acquire more connected features by using a similar way in Sect.~\ref{sect:finger_core_extraction}. }





\textbf{Limitation of using a static top layer: } 
Although previous works \cite{aldrich2016viscous, lukasczyk2017viscous, luciani2018details} and this paper use a static top layer (Sect.~\ref{sect:segmentation}) and acquire good results, physically, the top layer initially grows diffusively in time, i.e., as the square root of time, which is much slower than the convective time scale over which the fingers grow. Once the instability is triggered, this diffusive boundary layer becomes thin. Since the top layer is varying in nature, one limitation of using a static top layer is which may not accurately segment the structures near finger roots. 




\textbf{Scalability of the tracking graph: } 
\remark{When we use a screen resolution of $2880 \times 1800$, the tracking graph can display four consecutive timesteps, and show around one hundred fingers and hundreds of branches for each timestep.} 
If there are more than one hundred fingers at a timestep, the visual clutter becomes a bottleneck to arrange all the fingers in the same row. To remedy this scalability issue, a larger screen, more simplification of finger representations, and allowing multi-level exploration of fingers may be helpful. 




\textbf{Future works: } 
Our collaboration with the earth scientists continues to use the visualization and analysis tools developed in this work to study the scaling behavior associated with pattern formation in flow instabilities. Specifically, the topological measurements offered by our methods, including the number of branches and the number of critical points, can be studied for discovering new scaling laws. 



\section{Conclusion}
In this paper, to detect viscous and gravitational fingering, we present a novel geometry-driven approach with a ridge voxel detection guided finger core extraction, a finger skeletonization and pruning method, and a spanning tree based finger branch extraction. An interactive visual-analytics system with a novel geometric-glyph augmented tracking graph is established for scientists to track the geometric growth, merging, and splitting of fingers over time in detail. The earth scientists recognized the value of our work and thought that this could reduce their workload for the analysis of fingers both in space and time significantly.

\ifCLASSOPTIONcompsoc
  \section*{Acknowledgments}
\else
  \section*{Acknowledgment}
\fi
We thank Yujia Wang and Junpeng Wang for providing design suggestions on the geometric-glyph augmented tracking graph. Also, we thank Amin Amooie for the discussion on earth science. This work was supported in part by UT-Battelle LLC 4000159557, Los Alamos National Laboratory Contract 471415, and NSF grant SBE-1738502.


\ifCLASSOPTIONcaptionsoff
  \newpage
\fi



\bibliographystyle{IEEEtran}
\bibliography{finger}




%


\begin{IEEEbiography}[{\includegraphics[width=1in,height=1.25in,clip,keepaspectratio]{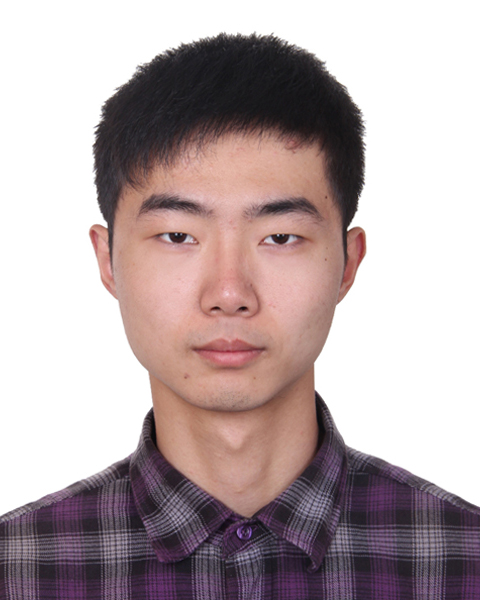}}]{Jiayi Xu} 
is a Ph.D. student in Computer Science and Engineering at the Ohio State University. He received his B.E. degree with Chu Kochen Honors from the College of Computer Science and Technology at Zhejiang University in 2014. His research interests include graph visualization and scientific feature tracking. 
\end{IEEEbiography}


\begin{IEEEbiography}[{\includegraphics[width=1in,height=1.25in,clip,keepaspectratio]{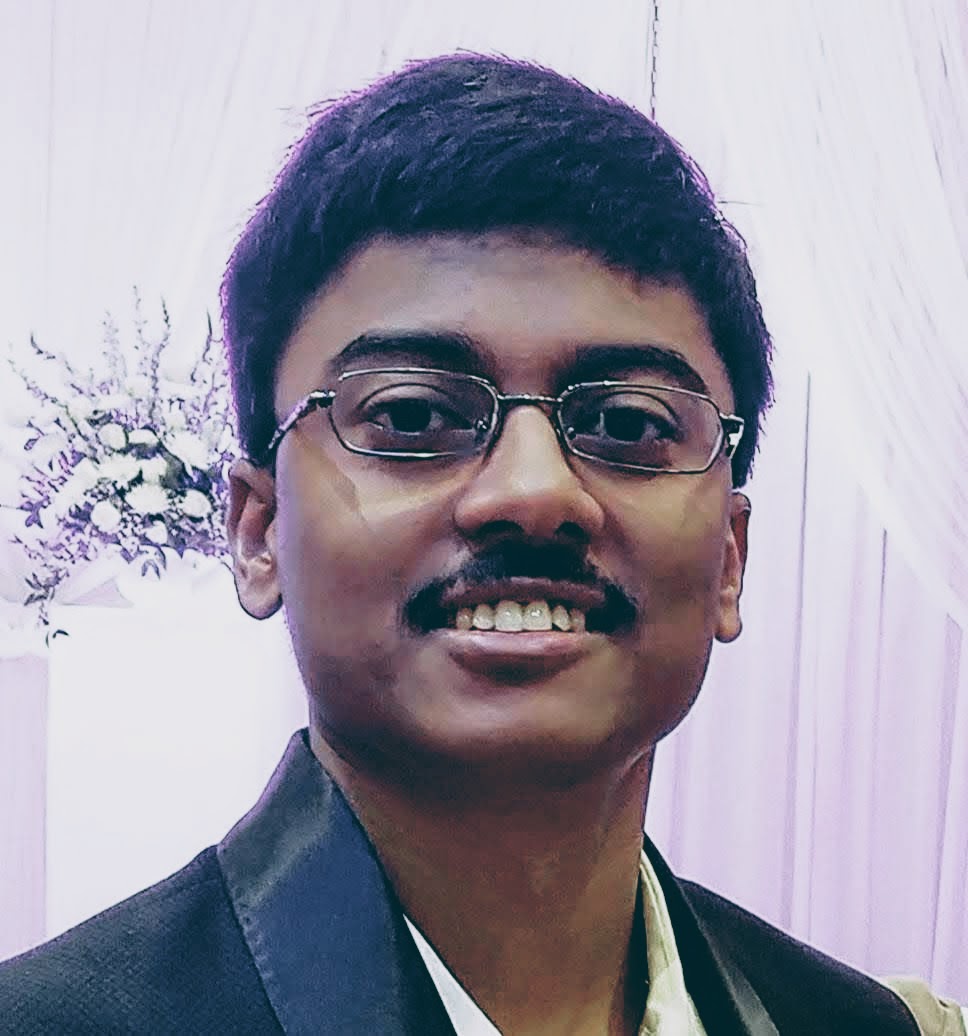}}]{Soumya Dutta} 
is a staff scientist in Data Science at Scale team at Los Alamos National Laboratory. He received his B.Tech. degree in Electronics and Communication Engineering from West Bengal University of Technology in August 2009, and M.S. and the Ph.D. degree in Computer Science and Engineering from the Ohio State University in May 2017 and May 2018 respectively. His research interests are statistical data summarization and analysis; in situ data analysis, reduction, and feature exploration; uncertainty analysis; and time-varying, multivariate data exploration.
\end{IEEEbiography}


\begin{IEEEbiography}[{\includegraphics[width=1in,height=1.25in,clip,keepaspectratio]{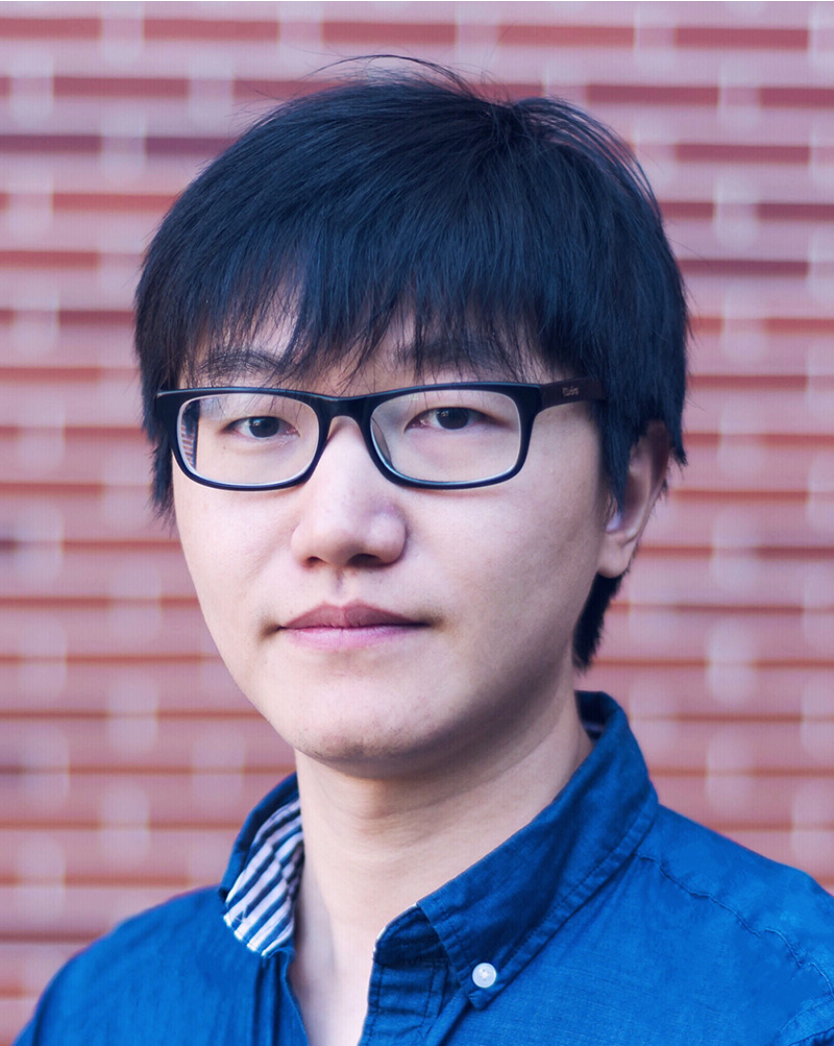}}]{Wenbin He}
is a Ph.D. student in computer science and engineering at the Ohio State University. He received his B.S. degree from the Department of Software Engineering at Beijing Institute of Technology in 2012. His research interests include visualization and analysis of large-scale scientific data, uncertainty visualization, and flow visualization.
\end{IEEEbiography}


\begin{IEEEbiography}[{\includegraphics[width=1in,height=1.25in,clip,keepaspectratio]{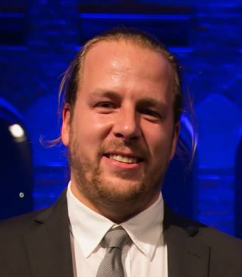}}]{Joachim Moortgat} 
is an Associate Professor in the School of Earth Sciences, Ohio State University. Moortgat holds MS degrees in theoretical physics and astrophysics, both from Utrecht University, and a PhD in astrophysics from the Radboud University, The Netherlands. Moortgat was the recipient of the 2014 SPE Cedric K. Ferguson Medal awarded by the Society of Petroleum Engineers. His research interests lie in the theory and advanced numerical modeling of compositional multiphase flow in subsurface fractured porous media. 
\end{IEEEbiography}


\begin{IEEEbiography}[{\includegraphics[width=1in,height=1.25in,clip,keepaspectratio]{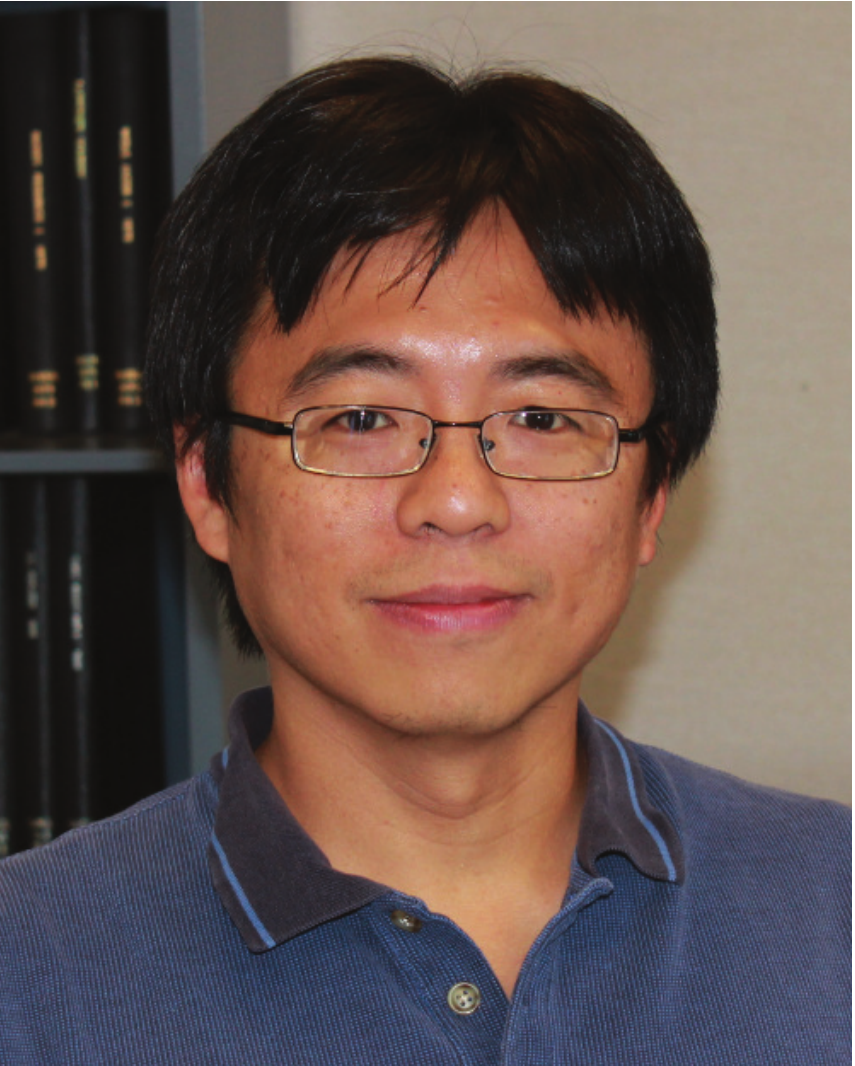}}]{Han-Wei Shen}
is a full professor at the Ohio State University. He received his B.S. degree from Department of Computer Science and Information Engineering at National Taiwan University in 1988, the M.S. degree in computer science from the State University of New York at Stony Brook in 1992, and the Ph.D. degree in computer science from the University of Utah in 1998. From 1996 to 1999, he was a research scientist at NASA Ames Research Center in Mountain View California. His primary research interests are scientific visualization and computer graphics. He is a winner of the National Science Foundation’s CAREER award and U.S. Department of Energy’s Early Career Principal Investigator Award. He also won the Outstanding Teaching award twice in the Department of Computer Science and Engineering at the Ohio State University.
\end{IEEEbiography}









\end{document}